 \documentclass[twocolumn]{aastex61}

\usepackage{amsmath}
\usepackage{amssymb}
\usepackage[utf8]{inputenc}
\usepackage{xspace}
\usepackage[british]{babel}
\usepackage{times}
\usepackage{graphicx}
\usepackage{tabulary}
\usepackage{color}
\usepackage{multirow}
\usepackage{float}
\usepackage{url}

\received{April 1, 2017}
\submitjournal{ApJ}

\newcommand{\ie}{i.\,e.\;}
\newcommand{\eg}{e.\,g.\;} 
\newcommand{\tv}{\theta_\mathrm{v}}
\newcommand{\tj}{\theta_\mathrm{jet}}
\newcommand{\lalinf}{\texttt{LALINFERENCE}\xspace}
\newcommand{\dL}{d_{\mathrm{L}}}
\newcommand{\Flim}{F_{\mathrm{lim}}}
\newcommand{\Sgnl}{\mathcal{S}}
\newcommand{\us}[1]{_{\mathrm{#1}}} 

\DeclareMathOperator*{\argmax}{arg\,max}

\newcommand{\simpropto}{\mathrel{\vcenter{
  \offinterlineskip\halign{\hfil$##$\cr
    \propto\cr\noalign{\kern2pt}\sim\cr\noalign{\kern-2pt}}}}}

\shorttitle{Where and when to look for EM counterparts}
\shortauthors{Salafia et al.}

\begin{document}

\title{Where and when: optimal scheduling of the electromagnetic follow-up of gravitational-wave events based on counterpart lightcurve models}

\correspondingauthor{Om Sharan Salafia}
\email{omsharan.salafia@brera.inaf.it, omsharan.salafia@gmail.com}

\author[0000-0003-4924-7322]{Om Sharan Salafia}
\affiliation{Universit\`a degli Studi di Milano-Bicocca, Piazza della Scienza 3, I-20126 Milano, Italy}
\affiliation{INAF - Osservatorio Astronomico di Brera Merate, via E. Bianchi 46, I–23807 Merate, Italy}
\affiliation{INFN - Sezione di Milano-Bicocca, Piazza della Scienza 3, I-20126 Milano, Italy}
\nocollaboration

\author[0000-0002-3370-6152]{Monica Colpi}
\affiliation{Universit\`a degli Studi di Milano-Bicocca, Piazza della Scienza 3, I-20126 Milano, Italy}
\affiliation{INFN - Sezione di Milano-Bicocca, Piazza della Scienza 3, I-20126 Milano, Italy}
\nocollaboration

\author[0000-0003-1643-0526]{Marica Branchesi}
\affiliation{Università degli Studi di Urbino ”Carlo Bo”, via A. Saffi 2, 61029, Urbino, Italy}
\affiliation{INFN, Sezione di Firenze, via G. Sansone 1, 50049, Sesto Fiorentino, Italy}
\nocollaboration

\author[0000-0003-3768-9908]{Eric Chassande-Mottin}
\affiliation{APC, Univ Paris Diderot, CNRS/IN2P3, CEA/Irfu, Obs de Paris, Sorbonne Paris Cité, France}
\nocollaboration

\author[0000-0001-5876-9259]{Giancarlo Ghirlanda}
\affiliation{INAF - Osservatorio Astronomico di Brera Merate, via E. Bianchi 46, I–23807 Merate, Italy}
\affiliation{Universit\`a degli Studi di Milano-Bicocca, Piazza della Scienza 3, I-20126 Milano, Italy}
\nocollaboration

\author[0000-0002-0037-1974]{Gabriele Ghisellini}
\affiliation{INAF - Osservatorio Astronomico di Brera Merate, via E. Bianchi 46, I–23807 Merate, Italy}
\nocollaboration

\author{Susanna Vergani}
\affiliation{GEPI, Observatoire de Paris, PSL Research University, CNRS, Univ. Paris Diderot, Sorbonne Paris Cité, Place Jules Janssen, 92195 Meudon, France}
\affiliation{Institut d’Astrophysique de Paris, Université Paris 6-CNRS, UMR7095, 98bis Boulevard Arago, 75014 Paris, France}
\affiliation{INAF - Osservatorio Astronomico di Brera Merate, via E. Bianchi 46, I–23807 Merate, Italy}
\nocollaboration

\begin{abstract}
The electromagnetic (EM) follow-up of a gravitational wave (GW) event requires to scan a wide sky region, defined by the so called ``skymap'', for the detection and identification of a transient counterpart. We propose a novel method that exploits information encoded in the GW signal to construct a ``detectability map'', which represents the time-dependent (``when'') probability to detect the transient at each position of the skymap (``where''). Focusing on the case of a neutron star binary inspiral, we model the associated short gamma-ray burst afterglow and macronova emission, using the probability distributions of binary parameters (sky position, distance, orbit inclination, mass ratio) extracted from the GW signal as inputs. The resulting family of possible lightcurves is the basis to construct the detectability map. As a practical example, we apply the method to a simulated GW signal produced by a neutron star merger at 75 Mpc whose localization uncertainty is very large ($\sim$ 1500 deg$^2$). We construct observing strategies based on the detectability maps for optical, infrared and radio facilities, taking VST, VISTA and MeerKAT as prototypes. Assuming limiting fluxes of $r\sim 24.5$, $J\sim 22.4$ (AB magnitudes) and $500\,\mathrm{\mu Jy}$ ($1.4\,\mathrm{GHz}$) for $\sim 1000$ s of exposure each, the afterglow and macronova emissions are successfully detected with a minimum observing time of 7, 15 and 5 hours respectively.

\end{abstract}

\keywords{gravitational waves -- stars: binaries --  gamma-ray burst: general -- stars: neutron -- methods: statistical}

\section{Introduction}

The first detection \citep{TheLIGOScientificCollaboration2016} of gravitational waves (GW hereafter) from the inspiral and merger of a black hole binary, followed by a second \citep{TheLIGOScientificCollaboration2016a} and a third one \citep{TheLIGOScientificCollaboration2017}, suddenly turned these fascinating, theoretical objects into real astronomical sources.

When such a compact binary coalescence is detected, analysis of the GW signal and comparison with carefully constructed templates of the waveform \citep{TheLIGOScientificCollaboration2016b,Abbott2016b} enables the extraction of precious information about the parameters of the binary and of the remnant. The identification of an electromagnetic (EM) counterpart would increase further the scientific outcome of the detection, e.g. by enabling the identification of the host galaxy, by providing hints about the environment surrounding the merger, by constraining theoretical models of EM counterparts and by reducing degeneracies in the GW extrinsic parameter space \citep{Pankow2016}. 

Several observatories, covering a large fraction of the EM spectrum, recently developed dedicated programs for the EM follow-up of GW events. The present main limitation for the detection of a possible EM counterpart is the large uncertainty on the sky localization of the GW source (see \eg \citealt{Singer2014,Berry2014}). The problem might be alleviated by targeting bright galaxies within the localization uncertainty region \citep{Nuttall2010,Abadie2012}, and the selection of target galaxies can also take into account the sky-position-conditional posterior distribution of the source luminosity distance \citep{Hanna2013,Nissanke2013,Gehrels2016,Singer2016}. The aim of this work is to propose an additional way to use information encoded in the GW signal to optimize the follow-up strategy for each single event, namely to combine posterior distributions of the compact binary parameters and available models of the EM emission to predict the best timing for the observation of different parts of the GW skymap. Such an approach can be applied in cases when a model of the expected EM counterpart is available, and it is especially useful when the lightcurve predicted by the model depends on (some of) the compact binary parameters. 

\subsection{The first electromagnetic follow-ups}
The observation campaigns that followed up the first detections of GWs were very extensive. Hundreds of square degrees within the GW sky localization were covered by wide-field telescopes \citep{Abbott2016a}. Target areas were selected in order to maximize the contained GW source posterior sky position probability, incorporating telescope visibility constraints \citep[\eg][]{Kasliwal2016}. In some cases, models of the expected EM counterpart emission were used to estimate the optimal search depth \citep[\eg][]{Soares-Santos2016}; other searches combined the posterior sky position probability map with the areal density and luminosity of nearby galaxies to select the best target fields \citep[\eg][]{Evans2016, Diaz2016}. Observations were concentrated during the first days after the events and repeated weeks to months later to search for both rapid and slowly evolving possible counterparts. 

\subsection{Candidate EM counterparts}

It is not clear whether an EM counterpart should be expected in the case of a binary of black holes (BH-BH), due to the unlikely presence of matter surrounding the binary (but see \citealt{Yamazaki2016,Perna2016,Loeb2016}); on the other hand, if the merger involves a black hole and a neutron star (BH-NS) or two neutron stars (NS-NS), there are solid reasons to believe that EM emission should take place. The most popular mechanisms for such an emission in both BH-NS and NS-NS cases include prompt (gamma-ray) and afterglow (panchromatic) emission from a short gamma-ray burst (SGRB) jet, and ``macronova'' (optical/infrared) emission from ejecta launched during and after the merger, powered by the decay of unstable heavy nuclei resulting from r-process nucleosynthesis taking place within the neutron-rich ejecta during the early expansion phase. 

Many other promising EM counterparts have been proposed, \eg the long lasting radio transient \citep{Nakar2011} arising from the deceleration of the dynamical ejecta due to interaction with the interstellar medium (ISM), the jet cocoon emission \citep{Lazzati2016,Gottlieb2017} or the spindown-powered emission described by \citet{Siegel2016} in the case when a (meta-)stable neutron star is left after the merger. To keep the discussion as simple as possible, in this paper we will only consider the (Optical and Radio) SGRB afterglow and the dynamical ejecta macronova as examples, leaving the possibility to apply the present approach to other EM counterparts to future works. 

\subsection{The SGRB afterglow}

The detectability of the SGRB prompt emission depends crucially on the jet viewing angle $\tv$, \ie the angle between the jet axis and our line of sight. If the viewing angle is larger than the jet half opening angle $\tj$ (in other words, if the jet points away from the Earth), the prompt emission flux received by an observer on Earth is severely suppressed \citep[\eg][]{Salafia2016} due to relativistic beaming (by the compactness argument, the bulk Lorentz factor in GRB jets must be comparable to or larger than one hundred -- \eg \citealt{Lithwick2001} -- and estimates based on observations are sometimes even larger than a thousand -- as in the short burst GRB090510, see \citealt{Ghirlanda2009,Ackermann2010}). Since the typical half opening angle $\tj$ is somewhere between $5^\circ$ and $ 15^\circ$ \citep[\eg][]{Berger2014}, the prompt emission goes undetected in the majority of cases (for an isotropic population, the probability that $\tv<15^\circ$ is less than 2 percent). Soon after producing the prompt emission, the jet starts interacting significantly with the ISM, and a shock develops \citep{Meszaros1996}. Electrons in the shocked ISM produce synchrotron radiation, giving rise to a fading afterglow \citep[observed fo the first time by Beppo-SAX,][]{Costa1997}. Since the consequent deceleration of the jet reduces the relativistic beaming, an off-axis observer (who missed the prompt emission) could in principle detect the afterglow before it fades \citep{rhoads-balloon97}: in this case, the afterglow is said to be \textit{orphan}. No convincing detection of such a transient has been claimed to date, consistently with predictions for current and past surveys \citep{Ghirlanda2015,Ghirlanda2014}, but future deep surveys (\eg MeerKAT in the Radio -- \citealt{Booth2009}, LSST in the Optical -- \citealt{Ivezic2008}, eROSITA in the X-rays -- \citealt{Merloni2012}) are anticipated to detect tens to thousands of such events per year. 

Given the large uncertainty on the expected rate of NS-NS and BH-NS detections by the aLIGO and Advanced Virgo facilities in the near future \citep{LIGOScientificCollaboration2010,Kim2015,Dominik2015,DeMink2016,TheLIGOScientificCollaboration2016c} and the rather low expected fraction of GW events with an associated SGRB jet pointing at the Earth \citep{Ghirlanda2016,Patricelli2016,Wanderman2014a,Metzger2011}, the inclusion of orphan afterglows as potential counterparts is of primary importance to test the SGRB-compact binary coalescence connection.

\subsection{The dynamical ejecta macronova}

Despite the idea dates back to almost twenty years ago \citep{Li1998}, the understanding of the possible macronova emission following a compact binary merger has been expanded relatively recently, as a result of the combined effort of researchers with expertise in a wide range of areas. A non-exhaustive list of the main contributions should include:
\begin{itemize} 
 \item numerical simulations of the merger dynamics (relativistic simulations by many groups using different approaches -- \eg \citealt{Dietrich2017,Ciolfi2017,Radice2016,Sekiguchi2016,Ruiz2016,Giacomazzo2015,Bauswein2015,Just2015,East2015,Sekiguchi2015,Wanajo2014,Kiuchi2014,Tanaka2013,Rezzolla2010} -- and non relativistic simulations, especially by Stephan Rosswog and collaborators -- \eg \citealt{Rosswog2013});
 \item studies to assess the efficiency of r-process nucleosynthesis and the consequent heating rate due to heavy element decay in the various ejecta \citep{Freiburghaus1999,Rosswog2000a,Korobkin2012,Wanajo2014,Lippuner2015,Hotokezaka2015a,Eichler2016,Rosswog2016a}; 
 \item atomic structure modeling which revealed the role of lanthanides in the ejecta opacity evolution \citep{Kasen2013};
 \item simulations including neutrino physics to model the neutrino-driven wind and the associated macronova \citep{Dessart2008,Martin2015,Perego2017};
\end{itemize}
  Results (especially for the dynamical ejecta) from various research groups begin to converge, and the dependence of the emission features on the parameters of the binary is in the process of being understood. Both analytical and numerical models capable to predict the lightcurve have been developed recently  \citep{Barnes2013,Grossman2013,Kawaguchi2016,Dietrich2016,Barnes2016,Rosswog2016a}. The emission from the dynamical ejecta is generally thought to be isotropic, which is an advantage with respect to the SGRB afterglow (which is instead beamed) from the point of view of the EM follow-up. The energy reservoir is the ejected mass $M\us{ej}$, which depends most prominently on the mass ratio $q=M_1/M_2$ of the binary and on the neutron star compactness, which in turn reflects the mass of the neutron star and its equation of state (EoS). Exciting claims of the detection of possible macronova signatures in the afterglows of few short GRBs \citep{Tanvir2013, Yang2015, Jin2015a, Jin2016} are in the process of being tested by intensive observational campaigns. All this makes the macronova emission an extremely interesting candidate EM counterpart. 

  \subsection{Outline of this work}
  
  In \S\ref{sec:sketch} we introduce the idea of a follow-up strategy as a collection of observations that partially fill a ``search volume'' (search sky area $\times$ typical transient duration), stressing that the GW ``skymap'' (the sky position probability density) gives information about \textit{where} to observe, but not about \textit{when}. In \S\ref{sec:a_priori_detectability_construction} we show how \textit{a priori} information about the EM counterpart can be used to quantitatively define how likely the detection of the EM emission is if the observation is performed at time $t$, thus providing some information about how to explore the temporal dimension of the ``search volume''. In \S\ref{sec:a_posteriori_idea} we suggest that the same approach can be extended to use \textit{a posteriori} information extracted from the GW signal, provided that we have a way to link the properties of the inspiral to those of the EM counterpart (as shown in \S\ref{sec:binary_counterpart_link}). In \S\ref{sec:sky_position_conditional_idea} we go one step further by introducing the idea that the information on the inspiral parameters that we can extract from the GW signal has a dependence on sky position, and thus the clues (that we obtain from the GW signal analysis) about when to observe can also depend on the sky position. In \S\ref{sec:mappediminchia} we introduce a method to extract such information from the ``posterior samples'' obtained from the analysis of a GW signal, and in \S\ref{sec:example} we apply it to a synthetic example to show how it optimizes the follow-up strategy. Finally, we discuss the results in \S\ref{sec:discussion} and we draw our conclusions in \S\ref{sec:conclusions}.

\section{Where and when to look}

\subsection{A sketch of the design of a follow-up observation strategy}\label{sec:sketch}

A short time after the detection of a compact binary coalescence signal, the LIGO Scientific Collaboration and Virgo Collaboration share information about the event with a network of astronomical facilities interested in the EM follow-up. The most fundamental piece of information for the follow-up is the so-called ``skymap'', \ie the posterior sky position probability density, which we denote as $P(\alpha \,|\, \Sgnl)$. It represents the probability per unit solid angle that the source is at sky position $\alpha$, say $\alpha=(\rm RA,\,Dec)$, given the GW signal $\Sgnl$ detected by the interferometers ($\Sgnl$ here represents all information contained in the strain amplitudes measured by all interferometers in the network). In what follows, we will most often call this probability density ``skymap probability''. Imagine the EM counterpart appears at the GW position right after the event and never turns off. Assume that it can be found by comparison with previously available images of the sky, and that it can be easily identified by its spectrum or by another method. An ultra-simplified sketch of the obvious follow-up strategy would then be the following:
\begin{enumerate}
 \item find the smallest sky area $A_\omega$ containing a large fraction $\omega$ (say $\omega=90\%$) of the skymap probability $P(\alpha\,|\, \Sgnl)$;
 \item divide such area into patches of size $A_{\rm{FoV}}$ corresponding to the field of view of the instrument;
 \item observe the patches in decreasing order of skymap probability\footnote{to keep the discussion as simple as possible, we are neglecting the limitations due to observing conditions.};
 \item for each patch:
 \begin{enumerate}
  \item identify the new sources by comparison with archive images;
  \item perform a set of operations, including \eg cross-matching with catalogues and spectral charachterization, to discard known variable sources and unrelated transients in order to identify the counterpart.
 \end{enumerate}
\end{enumerate}

The expected EM counterparts are transients, thus a first modification to the above sketch must take into account the time constraints coming from our \textit{a priori} knowledge of the transient features. If we have a physical or phenomenological model of the transient and we have some hint about the distribution of the parameters of such model, we can construct a prior probability $P(F(t)>\Flim)$ that the transient flux (in a chosen band) $F$ is above some limiting flux $\Flim$ at a given time $t$ after the GW event. Hereafter, we will call such quantity ``a priori detectability''. The probability of detecting the transient at time $t$ by observing a sky position $\alpha$ with an instrument with field of view $A\us{FoV}$ and limiting flux $\Flim$ is then 
\begin{equation}
P({\mathrm{det}}\,|\,t, \alpha,\mathrm{FoV}) \sim A\us{FoV}\, P(\alpha\,|\,\Sgnl)\times P(F(t)>\Flim)
\end{equation}
where we are assuming a relatively small field of view in order to consider $P(\alpha\,|\,\Sgnl)$ constant over its area. This is nothing more than saying that the best place to look for the transient is the point of maximum skymap probability, at the time of highest a priori detectability. The probability of detection decreases both moving away from the point of maximum skymap probability and observing at a time when $P(F(t)>\Flim)$ is smaller. 

\begin{figure}
 \includegraphics[width=\columnwidth]{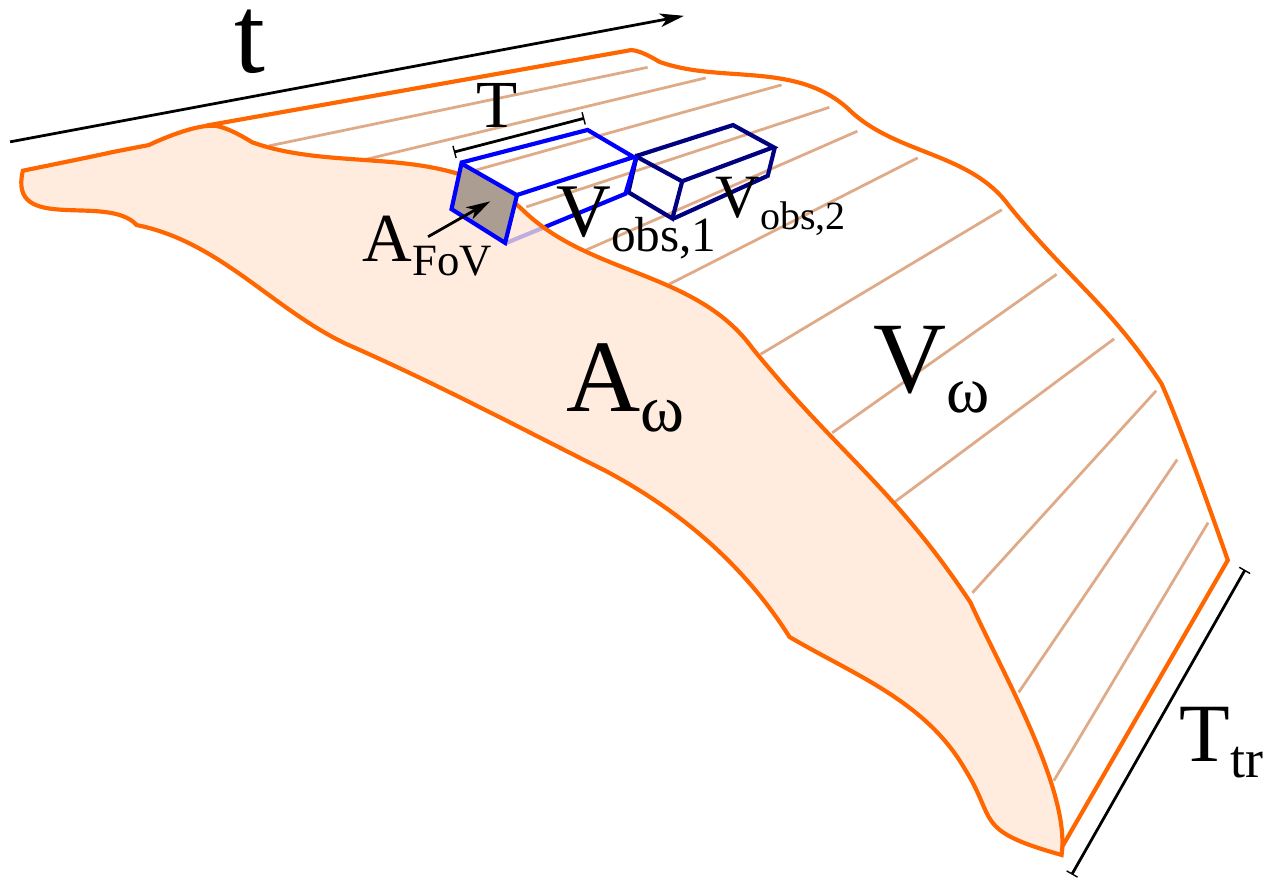}
 \caption{\label{fig:Vomega} Graphical representation of the follow-up strategy as a ``volume filling'' problem. $A\us{\omega}$ represents the sky region that contains a fraction $\omega$ of the sky position probability. The ``search volume'' is defined as the set of points $V_\omega = \left\lbrace (\alpha,t)\,|\,\alpha\in A_\omega\,\mathrm{and}\,t\in(0,T\us{tr})\right\rbrace$. Observations are sets that intersect the search volume, defined by a field of view $A\us{FoV}(\alpha\us{i})$ centered about sky position $\alpha\us{i}$, an exposure time $T$ and an observation time $t\us{obs,i}$ so that $V\us{obs,i}=\left\lbrace (\alpha,t)\,|\,\alpha\in A\us{FoV}(\alpha\us{i})\,\mathrm{and}\,t\in(t\us{obs,i},t\us{obs,i}+T)\right\rbrace$. Observations made by the same instrument cannot overlap on the time axis (unless the instrument can see more than one field at the same time). For the detection to be successful, the EM counterpart must be located within one of the $A\us{FoV}(\alpha\us{i})$ and its lightcurve must be above the detection threshold during the corresponding exposure time.}
\end{figure}

Let us work in the simplifying assumption that all observations have the same exposure $T$ and the same limiting flux $\Flim$. Let us denote by $T\us{tr}$ the most conservative (\ie largest) estimate of the transient duration, and let us define the ``search volume'' $V_\omega = A_\omega \times T\us{tr}$ (we refer to this set as a ``volume'' because it is 3-dimensional, even though the dimensions are $\mathrm{solid}\,\mathrm{angle}\times \mathrm{time}$ -- see Figure~\ref{fig:Vomega}). The follow-up strategy can then be thought of as an optimization problem, where one wants to (partially) fill the search volume $V_\omega$ with $N$ observations $V\us{obs, i} = A\us{FoV}(\alpha\us{i}) \times (t\us{obs, i},\, t\us{obs, i}+T)$ (where $\alpha\us{i}$ and $t\us{obs, i}$ are respectively the sky coordinates of the center of the field of view and the starting time of the $\mathrm{i}$-th observation) in order to maximise the detection probability $P(\mathrm{det}\,|\, \mathrm{strategy})$, which can be written as 
\begin{equation}
P(\mathrm{det}\,|\, \mathrm{strategy}) = \sum\us{i=1}^{N} \int_{A\us{FoV}(\alpha\us{i})} \mkern-45mu P(\alpha\,|\,\Sgnl)\,\mathrm{d}\alpha \times P(F(t\us{obs, i})>\Flim)
\end{equation}
with the constraint $N T \leq T\us{tr}$ (see Figure~\ref{fig:Vomega}).

The above paragraphs are essentially a formal description of the most basic follow-up strategy one can think of, which can be reduced to the principle ``try to arrange the observations in order to cover the largest possible fraction of the GW skymap around the time when the flux is expected to be high enough for a detection''. In this approach, the proper construction of the a priori detectability $P(F(t)>\Flim)$ is the key: it defines the time span within which the observations are to be performed, while the posterior sky position probability density defines the search area.


\subsubsection{How to construct the a priori detectability}
\label{sec:a_priori_detectability_construction}
In order to construct the a priori detectability $P(F(t) > \Flim)$, one must assume some prior probability density of the model parameters. Let us consider a simple, illustrative example. First, we construct a synthetic population of NS-NS inspirals whose properties roughly reproduce those expected for the population detected by Advanced LIGO; then we associate to each of them a jet afterglow and a macronova, under some assumptions. The detectable fraction of lightcurves in a given band, at a given time, will then constitute our estimate of the a priori detectability for this particular case. 
For the jet afterglow, we assume\footnote{these are typical reference values for short GRBs, though they suffer from the still limited number of reliable measurements available.} that all SGRB jets have an isotropic kinetic energy $E\us{K} = 10^{50}$ erg and a half-opening angle $\tj = 0.2$ radians (11.5 deg), and that they are surrounded by a relatively tenuous interstellar medium with constant number density $n\us{ISM} = 0.01$ cm$^{-3}$. We fix the microphysical parameters\footnote{We refer here to the standard synchrotron afterglow model, and we set the microphysical parameters $p=2.5$, $\epsilon_{\mathrm{e}}=0.1$ and $\epsilon_{\mathrm{B}}=0.01$. Such values, typical of Long GRBs \citep[\eg][]{Panaitescu2002,Ghisellini2009,Ghirlanda2015}, seem to be representative for SGRBs as well \citep{Fong2015}, despite the much smaller sample of broadband lightcurves available.} so that the only remaining parameters needed to predict the afterglow lightcurve of the SGRB are the distance $\dL$ and the viewing angle $\tv$. We will link the viewing angle to the binary orbit inclination, and the distance will be obviously set equal to that of the binary. Assuming two opposite jets launched perpendicular to the binary orbital plane, we have that
\begin{equation}
 \tv(\iota) = \left\lbrace\begin{array}{lcr}
        \iota & & 0\leq \iota<\pi/2\\
        \pi - \iota & & \pi/2 \leq \iota <\pi\\
       \end{array}\right.
\end{equation}
where $\iota$ is the angle between the normal to the orbital plane and the line of sight.

For the dynamical ejecta macronova, we evaluate the disk mass and the ejecta velocity using the fitting formulas of \citet{Dietrich2016}, and we use them as inputs to compute the lightcurve following \citet{Grossman2013} (using a constant grey opacity $\kappa = 10$ cm$^2$ g$^{-1}$), assuming a blackbody spectrum with effective temperature equal to that of the photosphere. The input compact binary parameters in this case are the masses $M_1$ and $M_2$. To determine the compactness and the baryon mass of the neutron stars, which are necessary to associate the dynamical ejecta mass $M\us{ej}$ and velocity v$\us{ej}$ to the merger through the fitting formulas of \citet{Dietrich2016}, we assume the H4 equation of state \citep{Lackey2006,Glendenning1991} which has a mid-range stiffness\footnote{This translates into mid-range values of the corresponding $M\us{ej}$ and v$\us{ej}$.} among those which are compatible with the observational constraints \citep{Ozel2016}.

First, we need to derive the proper distributions of distance and orbital plane inclination of the inspiral population detected by our interferometer network. For simplicity, we neglect the dependence of the network sensitivity on sky position and on the binary polarization angle $\psi$, and we assume that the maximum luminosity distance $d\us{L,max}$ out to which a NS-NS inspiral can be detected depends only on the binary plane inclination $\iota$ with respect to the line of sight, namely
\begin{equation}
 d\us{L,max}(\iota) = d\us{L,max}(0) \sqrt{\frac{1}{8}\left(1+6\cos^2\iota + \cos^4\iota\right)}
\end{equation}
where $d_\mathrm{L,max}(0)$ is the maximum luminosity distance out to which our network can detect a face-on inspiral. This expression accounts for the fact that gravitational radiation from a compact binary inspiral is anisotropic \citep{Schutz2011}. 
Assuming that NS-NS mergers are uniformly distributed in space and have isotropic orientations, their distance and inclination distributions are $P(d\us{L})\propto d\us{L}^2$ and $P(\iota)\propto \sin\iota$. By the above assumptions, the probability that a binary with luminosity distance $\dL$ and inclination $\iota$ is detected is
\begin{equation}
 P(\mathrm{det}\,|\,\dL,\iota)=\left\lbrace\begin{array}{lr}
                     1 & \mathrm{if}\;\dL<d_\mathrm{L,max}(\iota)\\
                     0 & \mathrm{otherwise}\\
                    \end{array}\right.
\label{eq:p_det}
\end{equation}
By Bayes' theorem, the probability distribution of distance and inclination of a detected NS-NS inspiral is then $P(\dL,\iota\,|\,\mathrm{det}) \propto P(\mathrm{det}\,|\,\dL,\iota)\times P(\dL) \times P(\iota)$, which gives
\begin{equation}
 P(\dL,\iota\,|\,\mathrm{det})\propto\left\lbrace\begin{array}{lr}
                     \dL^2\sin\iota & \mathrm{if}\;\dL<d_\mathrm{L,max}(\iota)\\
                     0 & \mathrm{otherwise}\\
                    \end{array}\right.
\label{eq:p_dl_iota}
\end{equation}
The corresponding probability distribution of inclination for detected inspirals (which is obtained by marginalisation of Eq.~\ref{eq:p_dl_iota} over $\dL$) is then the well known
\begin{equation}
 P(\iota\,|\,\mathrm{det}) = 7.6\times 10^{-2}\left(1+6\cos^2\iota + \cos^4\iota\right)^{3/2}\sin\iota
\end{equation}

For what concerns the distribution of masses $M_1$ and $M_2$, we simply assume a normal distribution with mean $1.35\,M_\odot$ and sigma $0.1\,M_\odot$ for both of them, which reproduces the mass distribution of known galactic NS-NS binaries \citep{Ozel2016}. 

Now, to construct the a priori detection probability $P(F(t) > \Flim)$ we adopt the following Monte Carlo approach: 
\begin{enumerate}
  \item we construct our synthetic population of $N$ inspirals sampling distances and inclinations from  $P(\dL,\iota\,|\,\mathrm{det})$ and masses $M_1$ and $M_2$ from the assumed normal distribution;
  \item we compute the flux $F\us{i} = F\us{A}(t,d\us{L,i},\tv(\iota\us{ i}))$ of the jet afterglow or $F\us{i} = F\us{M}(t,d\us{L,i},M\us{1,i},M\us{2,i})$ of the macronova in the chosen band for each sample;
  \item we estimate  $P(F(t) > \Flim)$ as the fraction of the $F\us{i}$'s that exceed $\Flim$.
\end{enumerate}

\begin{figure}
 \includegraphics[width=\columnwidth]{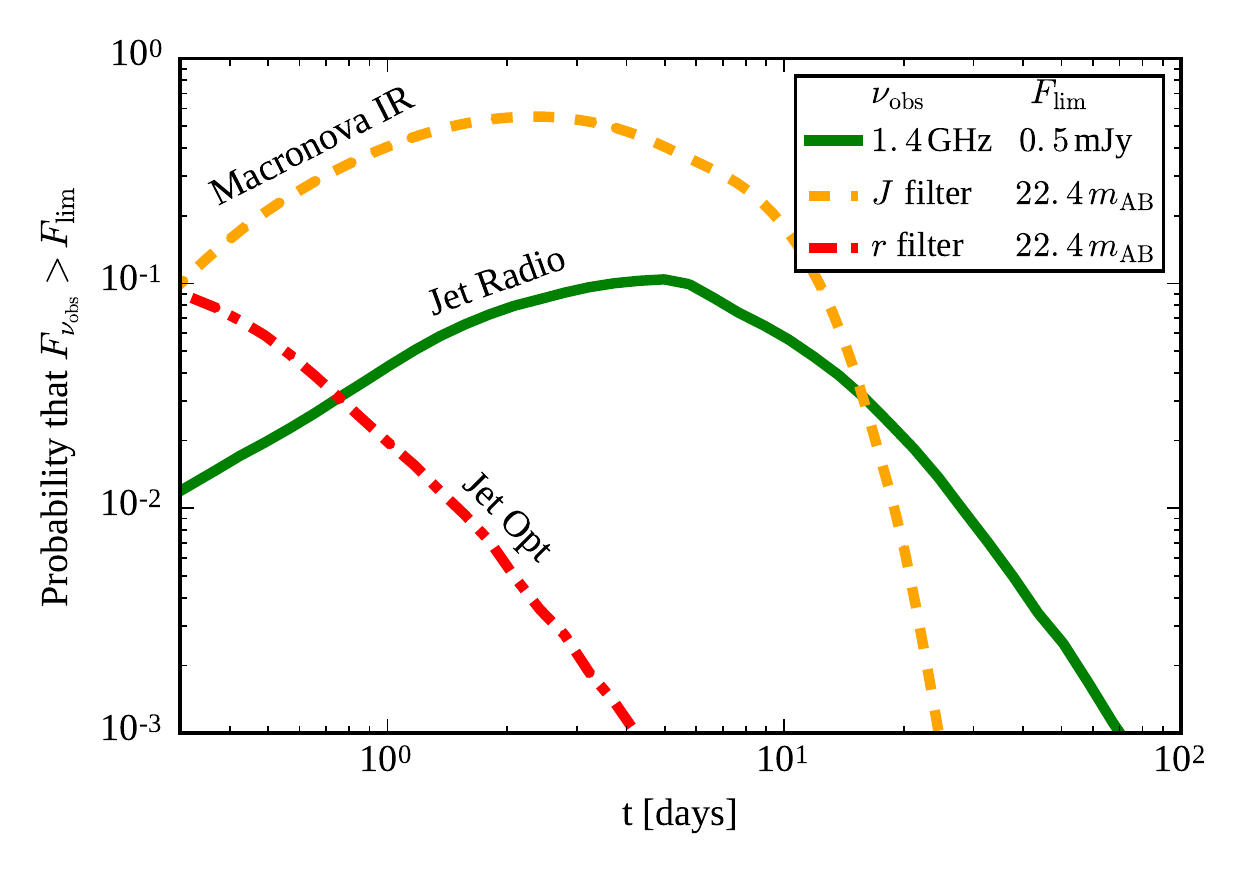}
 \caption{\label{fig:PF_gt_Flim_AP}``A priori detectability'', \ie a priori probability that the EM counterpart of a compact binary inspiral is detected if the observation is performed at a time $t$ after the merger, for observations in Radio at 1.4 GHz, in Infrared (IR) in the $J$ band, and in Optical in the \textit{r} band, with limiting fluxes of 0.5 mJy in Radio and 22.4 AB magnitude in IR and Optical. The Radio and Optical probabilities account only for the jet afterglow, while the IR probability accounts only for the dynamical ejecta macronova. Based on a series of simplifying assumptions, see \S\ref{sec:a_priori_detectability_construction}. 
 }
\end{figure}

Figure~\ref{fig:PF_gt_Flim_AP} shows the a priori detectability computed with the above method for $d_{\mathrm{L,max}}(0)=100$ Mpc \citep[which corresponds roughly to the sky-position averaged aLIGO range for an optimally oriented NS-NS inspiral with the O1 sensitivity, see][]{Abbott2016} for Radio, Infrared and Optical observations (see the caption for details). The sensitivity of Optical observations has been chosen to match the limiting magnitude of the VST follow-up of GW150914, see \citealt{Abbott2016a}. The flux of the jet afterglow has been computed using \texttt{BOXFIT} v. 1.0 \citep{VanEerten2011}.

A more accurate a priori detectability would require us to use astrophysically motivated priors on the other model parameters, such as the kinetic energy $E\us{K}$, the ISM number density $n\us{ISM}$, etc.
Moreover, the actual intrinsic mass distribution of neutron stars that merge within the frequency band of GW detectors might differ significantly from the assumed one. The curves shown in Fig.~\ref{fig:PF_gt_Flim_AP}, thus, must be taken as illustrative. 

The definitely higher detectability of the macronova is due to the fact that its emission is assumed to be isotropic, while the jet is fainter for off-axis observers.

\subsection{Two steps further: how to improve the strategy using posterior information about other parameters of the binary}\label{sec:a_posteriori_idea}

\subsubsection{The full posterior probability density in parameter space}

Parameter estimation techniques applied to a compact binary coalescence signal $\Sgnl$ result in a posterior probability density $P(\xi\,|\,\Sgnl)$, where $\xi \in \mathbb{R}^n$ is a point in the $n$-dimensional parameter space. The ``skymap probability'', \ie the posterior sky position probability density $P(\alpha\,|\,\Sgnl)$, is essentially the $P(\xi\,|\,\Sgnl)$ marginalised over all parameters but the sky position. Much more information is contained in the full posterior probability density, though, and some of it can be used to improve the design of the EM follow-up strategy.

\subsubsection{Relevant parameters in our case}\label{sec:binary_counterpart_link}

\begin{figure*}
 \includegraphics[width=\textwidth]{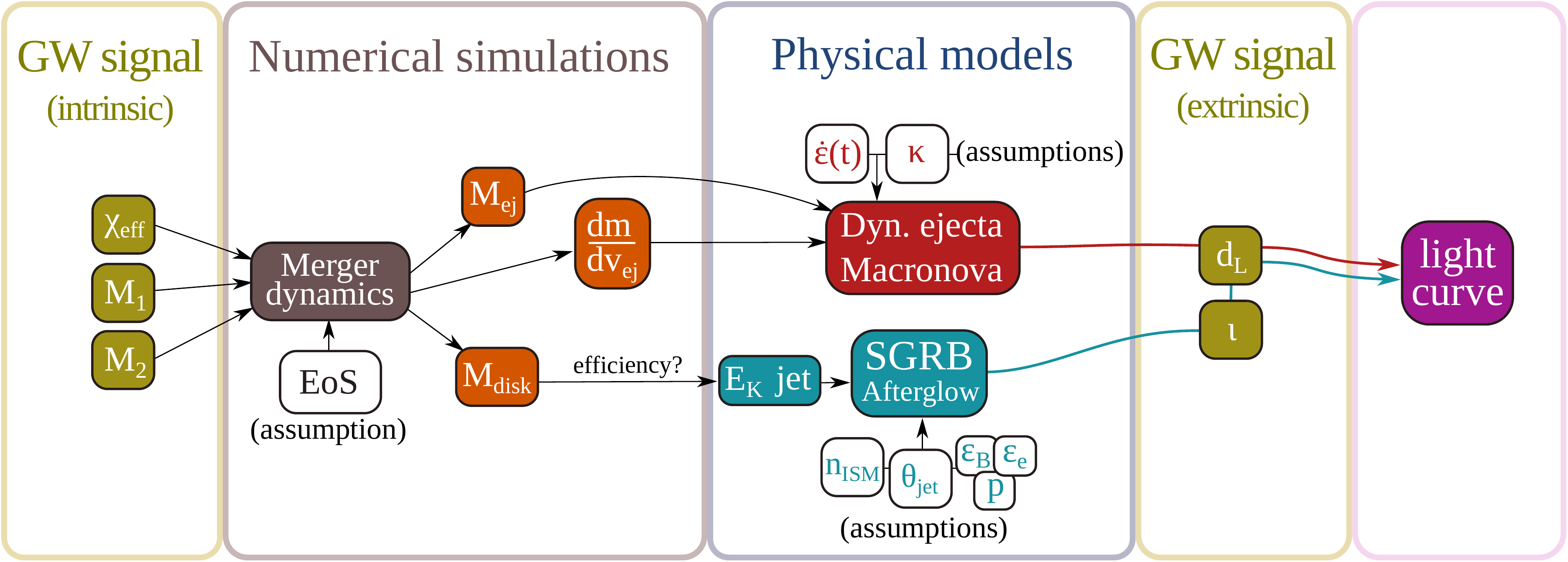}
 \caption{\label{fig:param_dependence}Sketch of the dependence of some parameters of the SGRB and dynamical ejecta macronova on the progenitor binary parameters. The masses and spin of the binary components, through the dynamics and assuming an equation of state (EoS), determine the masses $M\us{ej}$ \citep{Kawaguchi2016,Dietrich2016} and $M\us{disk}$ \citep{Foucart2012,Giacomazzo2013} of the dynamical ejecta and the accretion disk on the remnant compact object, respectively. The dynamics also determine the velocity profile $\mathrm{dm/dv\us{ej}}$ of the ejecta. Accretion on the remnant converts disk rest mass (with some unknown efficiency) into jet kinetic energy $E\us{K}$ \citep[\eg][]{Giacomazzo2013}, which constitutes the energy reservoir of the SGRB afterglow. By making assumptions on the remaining parameters, the lightcurves of the macronova and SGRB afterglow (possibly) associated to the merger can be predicted, taking into account the luminosity distance $d\us{L}$ and the inclination $\iota$ of the binary.}
\end{figure*}

In \S\ref{sec:a_priori_detectability_construction} we already made use of two extrinsic parameters of the compact binary inspiral which are relevant for the SGRB afterglow and the macronova, namely the luminosity distance $\dL$ and the binary inclination\footnote{the inclination of the orbital plane with respect to the line of sight can also be relevant for the neutrino-driven wind \citep{Martin2015} and the disk wind \citep{Kasen2014} macronovas, due to their axial geometry and to the possibility that the dynamical ejecta act as a ``lanthanide curtain'' obscuring their optical emission \citep{Rosswog2016a} if the binary is observed edge-on.} $\iota$. 

Recent works based on numerical simulations of NS-NS and BH-NS mergers \citep[\eg][]{Foucart2012,Giacomazzo2013,Hotokezaka2015,Kawaguchi2016,Dietrich2016} seem to indicate that the amount of matter in the remnant disk and in the dynamical ejecta, plus some other properties of the latter such as the velocity profile, depend in a quite simple way (once an equation of state is assumed) on the parameters of the binary prior to the merger, especially the masses $M_1$ and $M_2$ and the effective spin $\chi_{\rm eff}$ of the black hole in the BH-NS case. Such information can be used to predict the observed lightcurve of the associated macronovas \citep[\eg][]{Kawaguchi2016,Dietrich2016} and, with greater uncertainty, the energy in the GRB jet \citep[as in ][]{Giacomazzo2013}.

Summarizing, at least the following compact binary coalescence parameters are relevant in order to predict the lightcurve of the SGRB and/or of the dynamical ejecta macronova associated to the merger:
\begin{enumerate}
 \item the luminosity distance $\dL$ and the associated redshift $z$;
 \item the orbital plane inclination $\iota$ with respect to the line of sight;
 \item the component masses $M_1$ and $M_2$;
 \item the effective spin $\chi\us{eff}$ of the black hole in the BH-NS case.
\end{enumerate}

Figure~\ref{fig:param_dependence} represents a sketch of how the above parameters influence the properties of the SGRB afterglow and the dynamical ejecta macronova associated to the merger. The same approach can be adopted to link the properties of other EM counterparts (such as the long lasting radio transient described by \citealt{Nakar2011} or the X-ray spindown-powered transient described by \citealt{Siegel2016}) to those of the binary, whose distributions can be constrained by the GW signal.

In \S\ref{sec:example}, we will use the fitting formulas provided in \citet{Dietrich2016} to compute the posterior ejecta mass distribution associated to the example NS-NS inspiral treated in that section. We will refrain from deriving the SGRB jet energy from disk mass as suggested in Fig.~\ref{fig:param_dependence}, though, because that would require a detailed discussion about the proper disk mass energy conversion efficiency to be used, which is outside the scope of this work. We leave such a discussion to a future work. 

\subsubsection{A posteriori detectability}

In \S\ref{sec:sketch} we introduced the idea of ``a priori detectability'' $P(F(t) > \Flim)$, which can be regarded as the basic tool to set the timing of observations for the EM follow-up if no specific information about the source is available. Once the GW signal $\Sgnl$ is observed, information it carries can be used to construct a \textit{posterior} probability $P(F(t) > \Flim\,|\,\Sgnl)$ to better plan such observations. We will call it ``a posteriori detectablity''. If the \textit{a priori} detectability $P(F(t) > \Flim)$ is constructed using the \textit{prior} distributions of the parameters of the EM transient model, the \textit{a posteriori} detectability $P(F(t) > \Flim\,|\,\Sgnl)$ is obtained exactly the same way (as exemplified in \S\ref{sec:a_priori_detectability_construction}), but using the \textit{posterior} distributions of the relevant parameters.

\subsubsection{Detectability maps}\label{sec:sky_position_conditional_idea}

Several parameters of a compact binary inspiral are degenerate to some degree, \ie the same signal $\Sgnl$ can be produced by different combinations of the parameter values. These combinations, though, are not just uniformly distributed in some subset of the parameter space, but rather they follow fundamental relations which depend both on the nature of the source (the binary inspiral) and on the properties of the detector network (the locations and orientations of the interferometers, their antenna patterns, the noise power spectrum). In particular, distance, inclination, polarization angle, chirp mass and sky position of the binary share a certain degree of degeneracy: the same signal $\Sgnl$ can be produced by different combinations of values of these parameters and different realizations of the detector noises, which is the obvious reason why the sky position uncertainty is so large. For this reason, if we restrict the posterior probability density in parameter space to a certain point of the skymap, \ie we take the \textit{sky-position-conditional} posterior distribution of the physical parameters of the binary, in principle it will \textit{depend} on the chosen sky-position. 
Knowing the sky-position-conditional posterior probability distribution $P(\dL,\iota,M_1,M_2,...\,|\,\alpha, \Sgnl)$ of the relevant binary parameters at sky position $\alpha$, we can thus derive the corresponding distribution of the properties of the EM counterpart at that particular sky position, which means that we can construct a sky-position-conditional posterior detectability $P(F(t) > \Flim\,|\,\alpha,\Sgnl)$ which can be used as the basis of the EM follow-up strategy. We call this quantity ``detectability map''.

\subsection{Recap}

It is useful to summarize here the steps of increasing complexity that led us to the definition of the detectability maps:
\begin{itemize}
 \item We started by assuming an unrealistic model of the EM counterpart: a source that turns on at the GW time and never turns off. In this case, no timing information is needed for the follow-up strategy, which simply consist of scanning the localization uncertainty area, starting from the most probable sky location, until the source is found;
 \item If a model of the counterpart is available and prior distributions of the model parameters can be assumed (based on available astrophysical data or on an educated guess), the ``a priori detectability'' $P(F(t)>\Flim)$ can be constructed, as shown in \S\ref{sec:a_priori_detectability_construction}. This is the best follow-up timing information that can be constructed based on a priori knowledge only;
 \item After a GW signal $\Sgnl$ is detected and parameter estimation has been performed, prior distributions of the model parameters can be (partly) replaced with posterior distributions derived from the signal: the ``a posteriori detectability'' $P(F(t)>\Flim\,|\,\Sgnl)$ can be constructed. This exploits information contained in the GW signal, but it is still independent of the sky position;
 \item If the counterpart is \textit{assumed} to be located at a certain sky position, the corresponding sky-position-conditional posterior distributions can be used in place of the full posterior distributions. Indeed, given a signal $\Sgnl$, compact binary inspiral parameters compatible with $\Sgnl$ and a particular sky position are in general different from those compatible with $\Sgnl$ and another sky postion. By varying the assumed sky position on a grid that covers the whole skymap, one can then construct the ``detectability map'' $P(F(t)>\Flim\,|\,\alpha,\Sgnl)$.
\end{itemize}

Let us now introduce a method to compute the detectability maps and apply it to a practical example.

\section{How to construct and how to use the detectability maps}\label{sec:mappediminchia}

\subsection{Extraction of the sky-position-conditional posterior distributions using a simple method based on ``inverse distance weighting''}\label{sec:idw_based_method}

The extraction of the sky-position-conditional posterior distribution requires some multi-dimensional kernel density estimation (KDE) technique, to be applied to the posterior samples obtained from a parameter estimation pipeline run on the gravitational wave signals recorded by the detectors. Since the aim of this work is to propose a new approach in the design of the EM follow-up, rather than to discuss the technical subtleties of such multi-dimensional KDE, we adopt the following simple and intuitive method, which can be replaced with a more accurate one in a possible application of our approach to a real case.

Our simplified method to extract the sky-position-conditional posterior distribution $P(q\,|\,\alpha,\Sgnl)$ of a quantity $q$ at sky position $\alpha$ is based on the concept of ``inverse distance weighting'' \citep{Shepard1968}: we assume that each posterior sample $\left\lbrace\alpha\us{i},q\us{i},d\us{L,i},\iota\us{i},...\right\rbrace$ contributes to the $P(q\,|\,\alpha,\Sgnl)$ with a weight which is a decreasing function of the angular distance $\delta(\alpha,\alpha\us{i})$ between the posterior sample and the sky position $\alpha$. In particular, we assign a Gaussian weight to each posterior sample
\begin{equation}
 w\us{i} \propto \exp\left[-\frac{1}{2}\left(\frac{\delta(\alpha,\alpha\us{i})}{\sigma(\alpha)}\right)^2\right]
\end{equation}
where the bandwidth $\sigma(\alpha)$ is taken as 
\begin{equation}
\sigma(\alpha) = \sqrt{\frac{\sum\us{i=1}^N \left[\delta(\alpha,\alpha\us{i})-\left<\delta\right>\right]^2}{N-1}}\times N^{-1/5}
\label{eq:bandwidth}
\end{equation}
where $\left<\delta\right>$ is the arithmetic mean of the $\delta(\alpha,\alpha\us{i})$. The normalization of the weights is given by $\sum\us{i=1}^N w\us{i} = 1$.

The ideas behind this method are simply that the closer the posterior sample is to sky-position $\alpha$, the more it contributes to the conditional posterior distribution at that sky position, and that the influence of the posterior sample decreases as a Gaussian with increasing angular distance. The choice of the bandwidth (Eq.~\ref{eq:bandwidth}) is just ``Silverman's rule of thumb'' \citep{Silverman1982} for Gaussian KDE in a one dimensional parameter space (namely, the angular distance space).

The mean of $q$ at sky position $\alpha$ is thus computed as
\begin{equation}
 \left<q\right>_\alpha = \sum\us{i=1}^N w\us{i}\, q\us{i}
\end{equation}
and similarly the variance
\begin{equation}
 \mathrm{Var}_\alpha(q) = \left(\sum\us{i=1}^N w\us{i}\, q\us{i}\right)^2 - \sum\us{i=1}^N w\us{i}\, q\us{i}^2
\end{equation}
More generally, the sky-position-conditional posterior distribution of $q$ at sky position $\alpha$ is approximated as 
\begin{equation}
 P(q\,|\,\alpha,\Sgnl) \sim \sum\us{i=1}^N w\us{i}\,\mathrm{K}\left(\frac{q-q\us{i}}{\sigma_q}\right)
\end{equation}
where $\mathrm{K}(x)$ is some kernel function, and $\sigma_q$ is its bandwidth.

We performed tests to show that the above method yields consistent results (see the appendix). As one might expect, the results are accurate in sky regions where the distribution of posterior samples is sufficiently dense.

\subsection{The detectability map}
By the above method, we can thus define the sky-position-conditional posterior detectability estimate (\ie the detectability map) as
\begin{equation}
P(F(t) > \Flim\,|\,\alpha,\Sgnl) =\sum\us{i=1}^N w\us{i} \int_{\Flim}^\infty \,\mathrm{K}\left(\frac{F-F\us{i}(t)}{\sigma_F}\right)\,\mathrm{d}F 
\end{equation}
where $F\us{i}(t)$ represents the flux (in the chosen band) at time $t$ of the lightcurve computed using the i-th posterior sample parameter values, $F\us{i}(t)=F(t,d\us{L,i},\iota\us{i},...)$. If we approximate the kernel functions with delta functions $K(x)\sim \delta(x)$, the expression becomes
\begin{equation}
P(F(t) > \Flim\,|\,\alpha,\Sgnl) \sim \sum\us{i=1}^N w_i\,\mathrm{H}(F\us{i}(t)-\Flim) 
\end{equation}
where $\mathrm{H}(x)$ is the Heaviside function.  

\subsection{The earliest, best and latest detection time maps}\label{sec:best_detection_time}
By the above method, information encoded in the GW signal is used to estimate the detectability of the EM counterpart at a given time $t$, if it is at a certain sky position $\alpha$. By setting a minimum required detectability $\lambda$ one can define (for each sky position $\alpha$) a time interval during which $P(F(t) > \Flim\,|\,\alpha,\Sgnl)\geq\lambda$. If the detectability map $P(F(t) > \Flim\,|\,\alpha,\Sgnl)$ never reaches $\lambda$ for a certain sky position, that position is ``hopeless'', \ie there is too little chance of detecting the EM counterpart if it is located there. The earliest and latest detection time maps are then defined respectively as
\begin{equation}
\left\lbrace\begin{array}{l}
  \displaystyle t\us{E,\lambda}(\alpha) = \inf \left\lbrace t\,|\,p(t,\alpha) \geq \lambda\right\rbrace\\
  \rule{0pt}{16pt}
  \displaystyle t\us{L,\lambda}(\alpha) = \sup \left\lbrace t\,|\,p(t,\alpha) \geq \lambda\right\rbrace\\
\end{array}\right.
\end{equation}
where we set $p(t,\alpha) \equiv P(F(t) > \Flim\,|\,\alpha,\Sgnl)$ for ease of reading. Irrespectively of $\lambda$, the best detection time map can be defined as
\begin{equation}
 t\us{B}(\alpha) = \argmax\left\lbrace p(t,\alpha)\right\rbrace
\end{equation}
These maps are the simplest piece of information about ``where and when to observe'' that can be constructed using the detectability map. A follow-up strategy should then try to arrange observations so that a field centered at $\alpha$ is observed at a time as close as possible to $t\us{B}(\alpha)$, and in any case not earlier than $t\us{E,\lambda}(\alpha)$ or later than $t\us{L,\lambda}(\alpha)$. Let us now introduce a simple algorithm to construct such a follow-up strategy, after which we will be able to show a practical example (\S\ref{sec:example}).

\subsection{A follow-up strategy construction algorithm}\label{sec:algorithm}

\begin{figure}
 \includegraphics[width=\columnwidth]{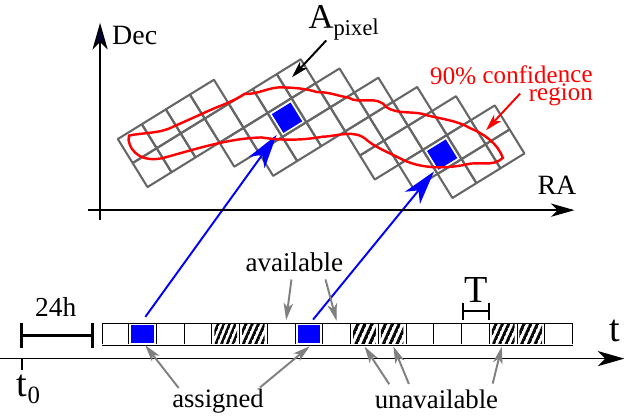}
 \caption{\label{fig:algorithm} Schematic representation of the strategy construction algorithm. At a given step, some of the available time slots have been already assigned, while others are still available. Observations that cover the current pixel are assigned to the available time slot corresponding to the maximum detectability.}
\end{figure}

In order to perform a first test of the approach outlined in the preceding sections, we will apply it to a simulated event and we will construct a ``simulated follow-up strategy'' based on it. To this end, we use an unambiguous algorithm to define the strategy for a given event and a given observing facility. To keep the discussion as simple as possible, we work in an idealized setting where all points of the skymap are observable by our facility during some pre-defined time windows. We assume that each observation covers an area $A\us{FoV}$ at observing frequency $\nu\us{obs}$ and that the limiting flux $F\us{lim}$ for detection is independent of sky position and is always reached after an integration time $T\us{int}$. The outline of the algorithm is the following:
\begin{itemize}
 \item we divide the skymap in patches, each representing a potential field to be observed;
 \item we define a list of available time slots (\ie possible observing time windows) on the time axis, starting 24 hours after the event (posterior samples are typically obtained after several hours or even days, so an earlier start would be unrealistic. Note that this does not mean that we discourage an earlier follow-up -- which is of great importance especially for the Optical and X-ray afterglow -- but only that this method may not be applicable with very low latency\footnote{In the example of \S\ref{sec:example}, actually, the only parameters needed to predict the possible SGRB afterglow light curves are $d\us{L}$ and $\iota$. Posterior distributions of these parameters (so called ``extrinsic'') can be obtained with a low latency analysis tool such as \texttt{BAYESTAR} \citep{Singer}.});
 \item starting from the patch with highest sky position probability, we check if the detectability map $P(F(t)>\Flim\,|\,\alpha,\Sgnl)$ within that patch exceeds $\lambda$ at some time within the available time slots: if it does not, we discard the patch (we choose not to observe it); if it does, we schedule the observations that cover that patch at the time when the detectability is highest, and we mark the corresponding time slots as not available anymore;
 \item we proceed to the next patch in descending order of skymap probability until the available time is over, or until all patches have been processed.
\end{itemize}
To keep the implementation of the above steps as simple as possible, we use a HEALPix tessellation of the sky \citep{Gorski2005} to define the observable fields: it is a way to divide the sky (\ie a sphere) into equal area patches, called ``pixels''. The ``order'' $N\us{side}$ of the HEALPix tessellation defines the number of pixels the sky is divided into, namely $N\us{pixels} = 12 N\us{side}^2$. The pixel area is then $A\us{pixel}=3438\,N\us{side}^{-2}$ deg$^2$. Thus we replace the actual observations of duration $T\us{int}$ and field of view $A\us{FoV}$ with ``pseudo observations'' of area $A\us{pixel}$ and effective duration $T=T\us{int} \times A\us{pixel}/A\us{FoV}$, choosing $N\us{side}$ in order to minimize the difference between $A\us{pixel}$ and $A\us{FoV}$. The algorithm is thus implemented as follows:
\begin{enumerate}
 \item we consider the posterior samples produced by a parameter estimation sampler (multiple sampling algorithms are implemented in LIGO's \texttt{LALINFERENCE} parameter estimation tool\footnote{\url{http://software.ligo.org/docs/lalsuite/lalinference/}}) applied to a simulated signal $\Sgnl$. These are points in the compact binary inspiral parameter space distributed according to the posterior probability density;
 \item we find the 90\% sky position confidence region of the source based on the posterior samples;
 \item we divide the sky into pixels according to a HEALPix tessellation of order $N\us{side}$, and we consider only those pixels which fall inside the 90\% sky position confidence region of the simulated signal;
 \item we associate to each pixel $p$ the integral of the skymap probability density $P_p=\int_{A\us{pixel}(p)}P(\alpha\,|\,\Sgnl)\,\mathrm{d}\Omega$ over the pixel area;
 \item we associate to each pixel $p$ the earliest and latest detection times $t\us{E,\lambda}(p)$ and $t\us{L,\lambda}(p)$ averaged over that pixel; pixels for which the detectability never reaches $\lambda$ are excluded from the list of possible observations;
 \item we divide the time axis into contiguous intervals (slots) of duration $T$ (the effective time needed to cover one pixel) and we mark some of the slots as ``available'' for the follow-up (based on the characteristics of the instrument);
 \item we sort the pixels in order of decreasing $P_p$ and, starting from the first ($p=1$) pixel, we do the following:
  \begin{enumerate}
    \item we check that at least one available time slot is comprised between $t\us{E,\lambda}(p)$ and $t\us{L,\lambda}(p)$: if not, no observation of the pixel is scheduled; otherwise, the available time slot where the detectability is maximum is assigned to the observation of the pixel;
  \item we proceed to the next pixel, until all available time slots are assigned, or until all pixels have been processed.
  \end{enumerate}
  
 \end{enumerate}
 
 Figure~\ref{fig:algorithm} shows a schematic representation of the algorithm described above.

 The output of the above algorithm is thus a list of observation times $t\us{obs}(p)$ that cover (part of) the skymap, giving priority to pixels with high skymap probability and trying to use the available time in a way that maximizes the probability to detect the transient. The inputs of the algorithm are:
 \begin{enumerate}
  \item the posterior sample list based on $\Sgnl$;
  \item the detectability threshold $\lambda$;
  \item the available time windows;
  \item the instrument observing frequency $\nu\us{obs}$, the field of view area $A\us{FoV}$, the limiting flux $\Flim$ and the corresponding integration time $T\us{int}$ (which should also include the slew time);
  \item the HEALPix tessellation order $N\us{side}$, which should give a pixel area $A\us{pixel}$ close to $A\us{FoV}$.
 \end{enumerate}
 
 The observations are given in order of decreasing ``importance'' (skymap probability). Once the list of observations is produced, one can decide to perform only the first $N$ observations that fit into the available telescope time. If all observations suggested by the algorithm can be performed within the available time, the excess time can be used \eg to take comparison images of some fields at different times (for the identification of transients or uncatalogued variable sources, in absence of previously available images) or to perform deeper observations for the characterization of the candidates.
 
 We are now ready to construct an example of how the use of sky-position-conditional posterior probabilities can help in the definition of an EM follow-up strategy.

\section{Test example: injection 28840, a NS-NS merger with an associated orphan afterglow}\label{sec:example}

\begin{figure}
 \includegraphics[width=\columnwidth]{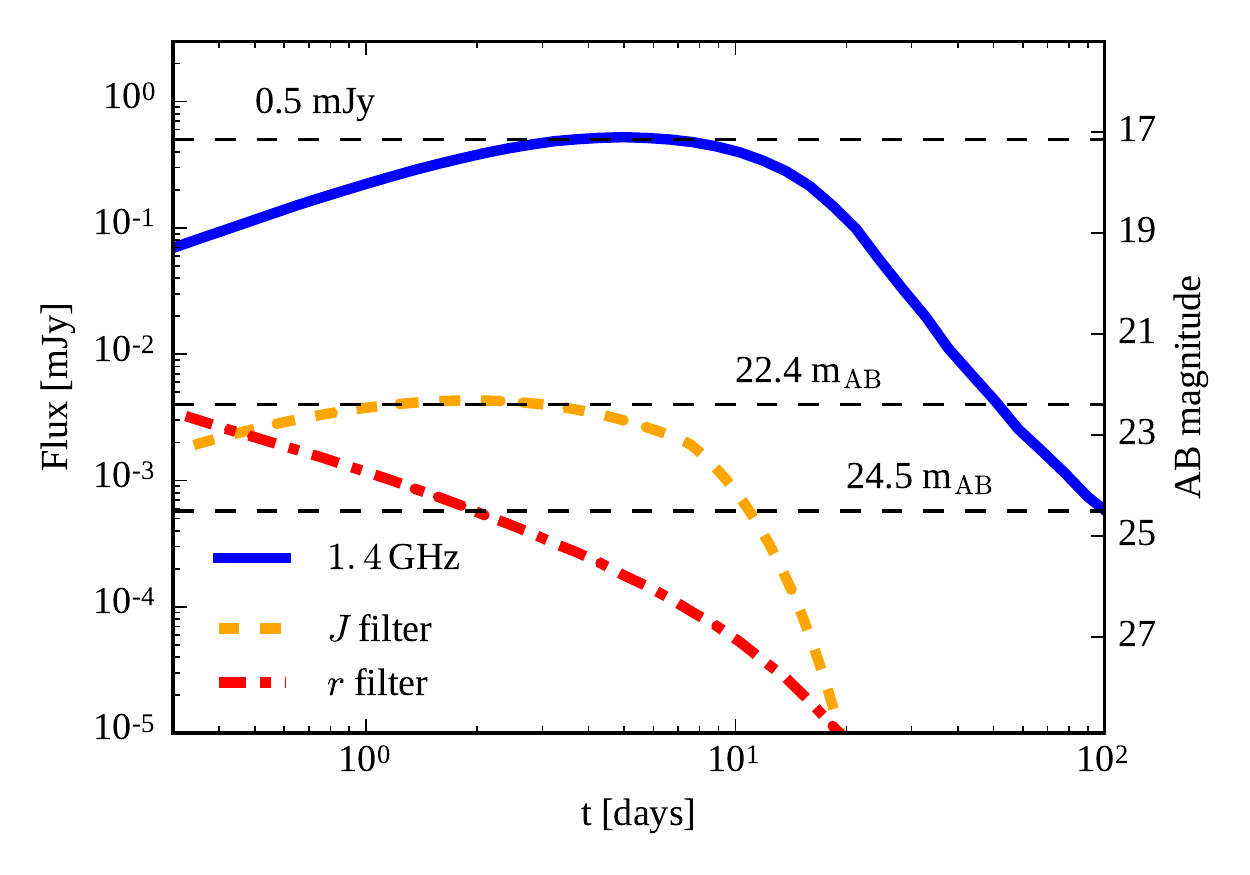}
 \caption{\label{fig:28840_lightcurves}Light curves of the EM counterparts associated to injection 28840 of F2Y. Jet afterglow: Radio (1.4 GHz -- blue line) and optical (\textit{r} filter -- red line); macronova: Infrared (\textit{J} filter -- yellow line). The same assumptions as in S\ref{sec:a_priori_detectability_construction} were adopted. The limiting fluxes for detection adopted in the test example (\S\ref{sec:example}) are shown with horizontal black dashed lines.}
\end{figure}

\begin{table*}
\caption{ Algorithm input parameters used to construct the example follow-up strategies.}
\label{tab:algorithm_inputs}
\setlength\extrarowheight{3pt}
\begin{tabular*}{\textwidth}{ccccccccc}
\hline
  Instrument & $\nu\us{obs}$ [Hz] & $\Flim$ [$\mu$Jy] & $A\us{FoV}$ [deg$^2$] & $T\us{int}$ [s] & $\lambda$ & $N\us{side}$ & Available time$^{\mathrm{a}}$ & Det. time$^{\mathrm{b}}$ [h] \\[1pt]
  \hline
  VST-like (shallow) & 4.8$\times 10^{14}$ (\textit{r} filter) & 4 (22.4 m$_\mathrm{AB}$) & 1 & 100 & 0.01 & 64 & 3 h/night & - \\
  VST-like (deep) & 4.8$\times 10^{14}$ (\textit{r} filter) & 0.58 (24.5 m$_\mathrm{AB}$) & 1 & 1000 & 0.05 & 64 & 3 h/night & 7 \\
  MeerKAT-like & 1.4$\times 10^{9}$ & 500 & 1.7 & 1000 & 0.05 & 64 & 20\% & 4.7 \\
  VISTA-like & 2.4$\times 10^{14}$ (\textit{J} filter) & 4 (22.4 m$_\mathrm{AB}$) & 1.5 & 1200 & 0.5 & 64 & 3 h/night & 14.5 \\
  \hline
 \end{tabular*}

\flushleft \footnotesize{$^{\mathrm{a}}$available observing time starts 24h after the event.\\$^{\mathrm{b}}$minimum observation time needed for detection.}
 
\end{table*}

As a test example, we consider injection number 28840 from the ``First two years of electromagnetic follow-up with Advanced LIGO and Virgo'' study \citep[][F2Y hereafter]{Singer2014}. The injection simulates the inspiral of a neutron star binary with $M_1=1.59 \,M_\odot$ and $M_2=1.53\,M_\odot$ at a luminosity distance $\dL = 75$ Mpc, with orbital plane inclination $\iota = 14^\circ$, at sky position $(\mathrm{RA}, \mathrm{Dec}) = (23^\mathrm{h}\,27^\mathrm{m}\,12^\mathrm{s}$, $-10^\circ\,30^\prime\,0^{\prime\prime})$, detected by the two-detector Advanced LIGO network on MJD 55483.27839 (\ie at 06:40:53 of the 14$^{\rm th}$ of October 2010 -- this is just a simulated event) adopting an early sensitivity curve corresponding to a binary NS range of 55 Mpc \citep{Barsotti2012}. Assuming that a relativistic jet with isotropic equivalent kinetic energy $E\us{K}=10^{50}$ erg and half-opening angle $\tj=0.2$ rad (which is less than the viewing angle, thus the afterglow is orphan) is launched perpendicular to the orbital plane right after the merger, we computed its afterglow lightcurve assuming an ISM number density $n\us{ISM} =0.01\,\mathrm{cm}^{-3}$, adopting the same microphysical parameters as in \S\ref{sec:a_priori_detectability_construction}, using \texttt{BOXFIT} v. 1.0 \citep{VanEerten2011}. The lightcurves at $\nu\us{obs}=1.4$ GHz and in the \textit{r} filter are shown in Figure~\ref{fig:28840_lightcurves}.

\begin{figure*}
 \includegraphics[width=\textwidth]{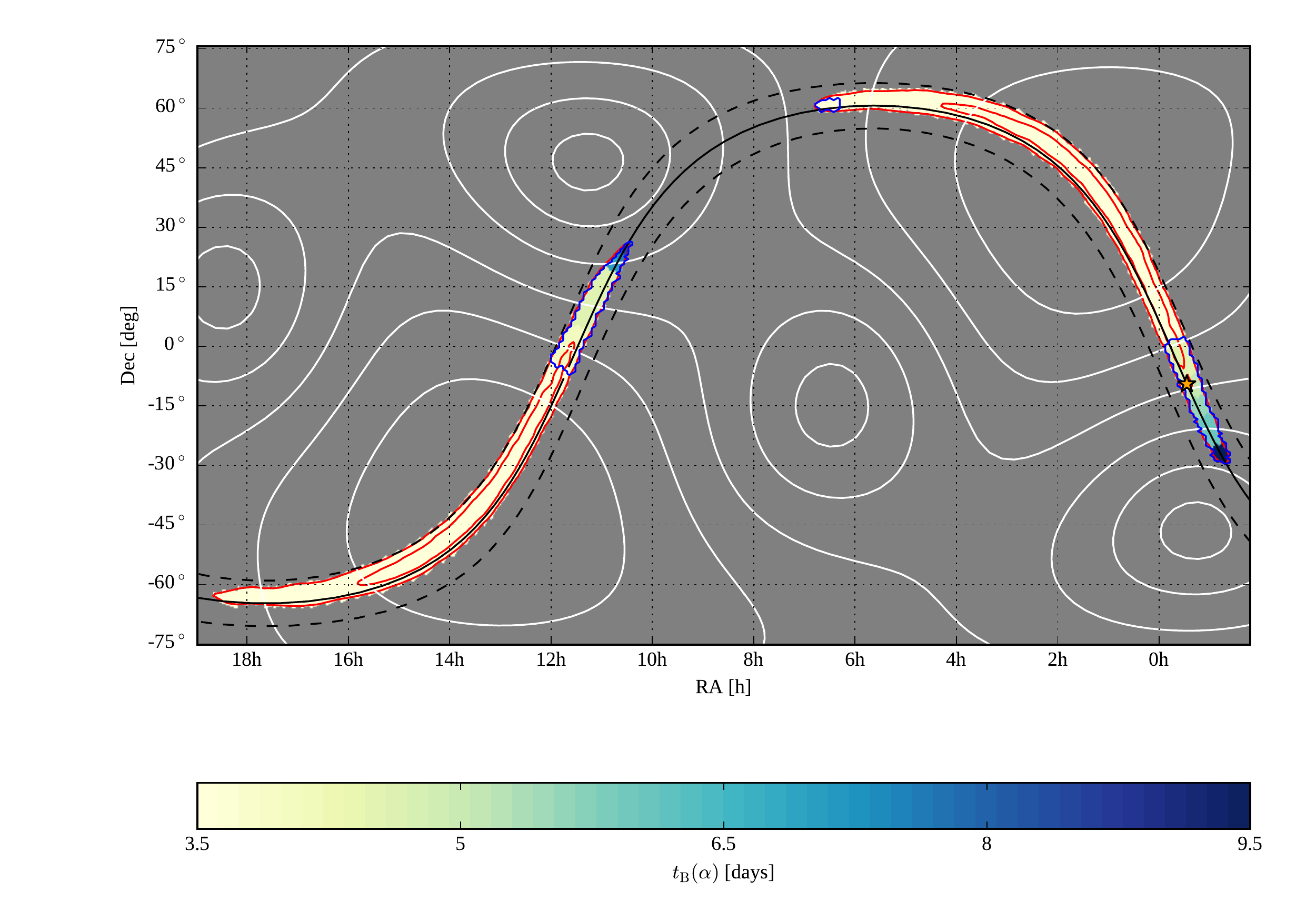}
 \caption{\label{fig:28840_skymap} Mercator projection of the best detection time map $t\us{B}(\alpha)$ at $\nu\us{obs}=1.4$ GHz for injection 28840 of F2Y. Red lines represent the contours of the 50 and 90 percent sky position confidence areas; the black solid line is the locus of sky positions that yield a GW signal arrival time delay of $\Delta t = 0.359$ ms (the ``true'' arrival time difference associated to the injection position) between the two LIGO detectors, and the black dashed lines correspond to $\Delta t \pm 1$ ms. The white contours are isolines of the average antenna pattern of the aLIGO network (the four smaller closed circular contours enclose the local and absolute minima, while the two large closed contours near the lower left and the upper right corners enclose the absolute maxima). The injection true position is marked with a star symbol. Regions marked by blue contours are those where the posterior detectability of the EM counterpart of our test example reaches the required 5 percent threshold.}
\end{figure*}

To produce our sky-position-conditional posterior distributions, we use 7962 posterior samples produced by one of the \lalinf parameter estimation samplers. Since we fixed the jet isotropic equivalent kinetic energy $E\us{K}=10^{50}$ erg, in this example the only binary parameters which are relevant to the posterior detectability of the jet afterglow are the luminosity distance and the binary plane inclination. Figure~\ref{fig:28840_skymap} shows the best detection time map at $\nu\us{obs}=1.4$ GHz produced using these samples (see the caption for additional information). The 90 percent sky position confidence area (represented by the larger red contours) covers approximately 1500 deg$^2$.

\subsection{Optical search}

\begin{figure}
 \includegraphics[width=\columnwidth]{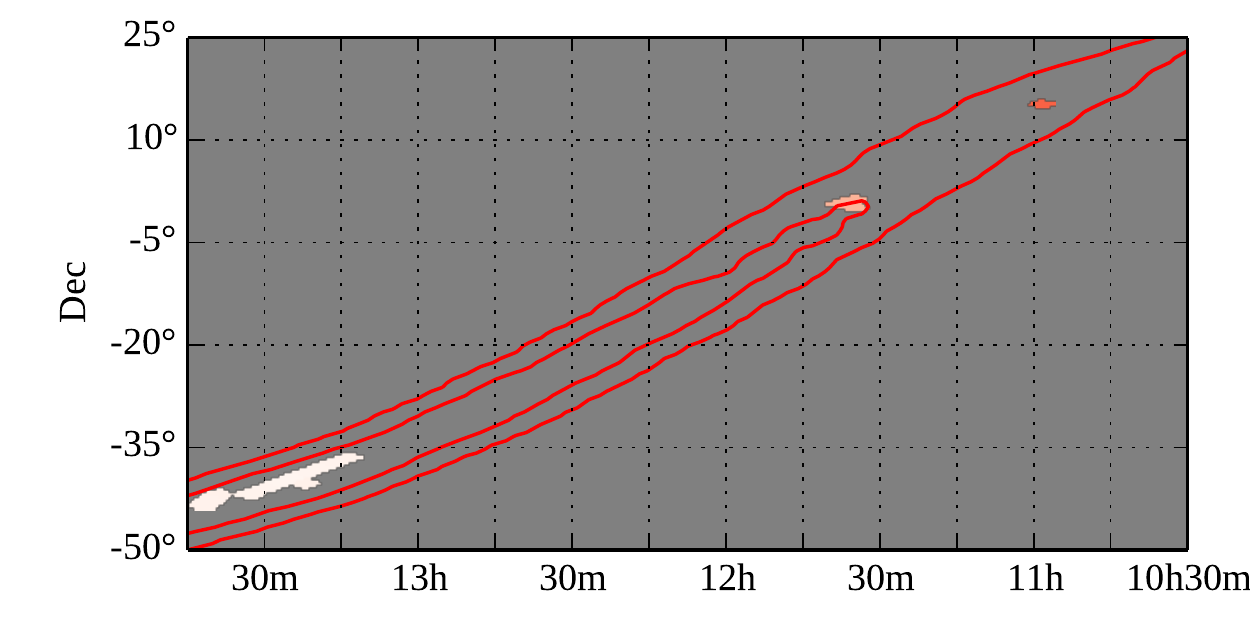}
 
 \includegraphics[width=\columnwidth]{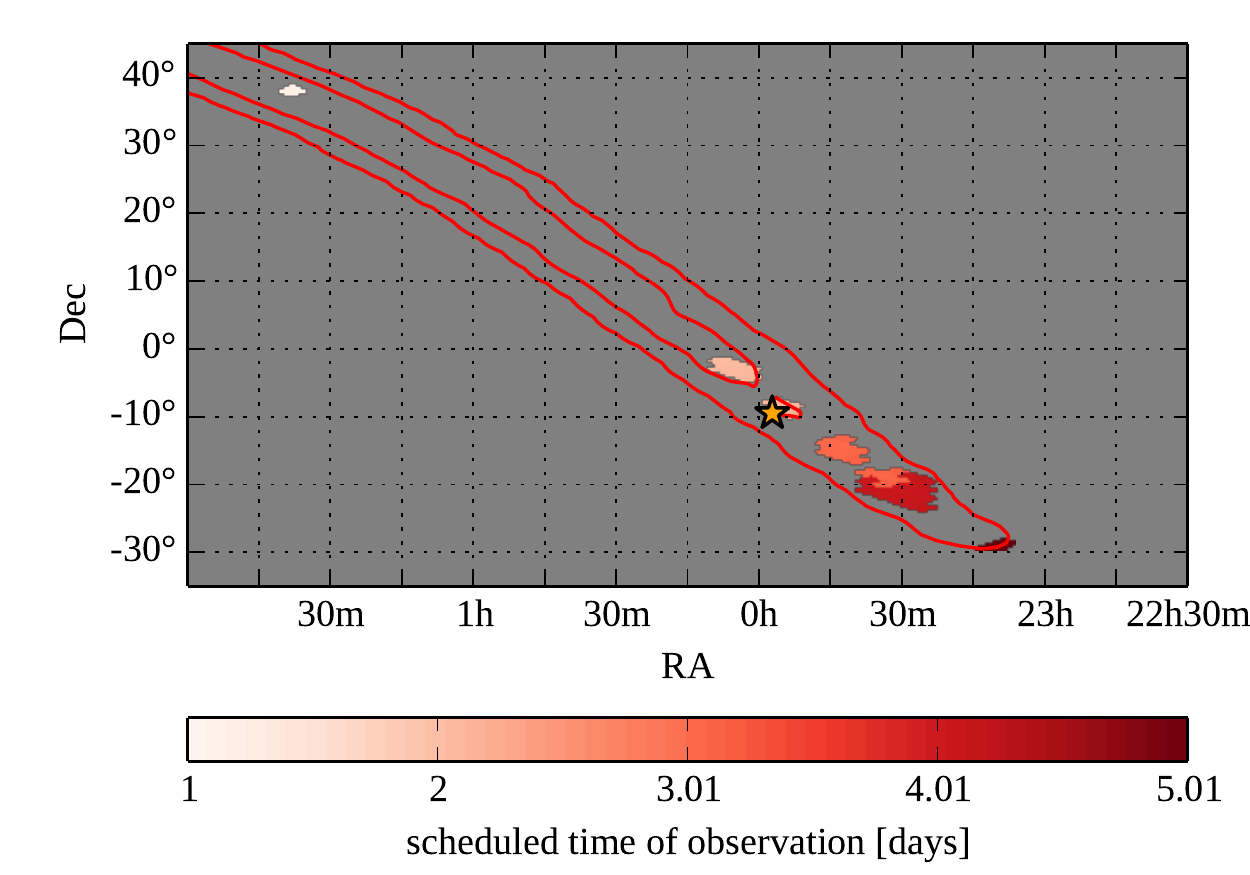}
 
 \caption{\label{fig:28840_obsmap_OPT} Time and position of the Optical follow-up observations of our test example (\S\ref{sec:example}). The star marks the injection position.}
\end{figure}

For our virtual optical EM follow-up, we adopt parameters inspired by the VST follow-up of GW150914 \citep{Abbott2016a}. We consider observations in the \textit{r} band ($\nu\us{obs} = 4.8\times 10^{14}$ Hz) with a detection limit $\Flim = 22.4$ AB magnitude, reached after $T\us{int} = 100$ seconds (slew + integration). The field of view is $A\us{FoV}= 1$ deg$^2$. With the adopted flux limit, the optical lightcurve (Fig.~\ref{fig:28840_lightcurves}) becomes too faint for detection after a few hours. Consistently, the detectability $P(F(t) > \Flim\,|\,\alpha,\Sgnl)$ is below 1 percent at all times $t>1$ d over the whole skymap except for a subregion of the skymap with a total area of 185 deg$^2$. We run the follow-up strategy construction algorithm allowing 3 hours of available time per night from day 1 to day 15. We set $N\us{side}=64$ which gives $A\us{pixel}=0.84$ deg$^2$ and $T=84$ s. Even adopting the very low detectability limit $\lambda=0.01$ (\ie allowing for observations with a detectability as low as 1 percent), only 24 ``pseudo observations'' (corresponding to 22 observations) are scheduled (all during the first available night), totalling 36 minutes of telescope time. The position of the EM counterpart is not contained in any of the observed fields.

The short integration time and the relatively shallow detection limit of the VST follow-up of GW150914 are good in order to cover the largest possible area during the next few nights after the event; if we wish to have some chance to detect a relatively dim optical afterglow like that in Fig.~\ref{fig:28840_lightcurves}, though, we need to go deeper, \ie we need longer integration times. We thus repeat the optical search with the same parameters as above, but with $T\us{int}=1000$ s and $\Flim=24.5$ m$\us{AB}$. We also raise the minimum detectability to $\lambda=0.05$, to avoid pointing fields with a very low detectability. The algorithm outputs 54 ``pseudo observations'', corresponding to 46 observations, totalling 15 hours of telescope time. The $p=25$ observation (\ie the 25th observation in descending order of $P_p$), scheduled $2\mathrm{d}\,2\mathrm{h}\,34\mathrm{m}\,45\mathrm{s}$ after the event, contains the EM counterpart. The flux at that time is 24.58 m$\us{AB}$, slightly dimmer than the required detection threshold, but detectable with a threshold S/N ratio of 5 under optimal observing conditions.

As explained in \S\ref{sec:algorithm}, the follow-up observations suggested by the algorithm are given in descending order of sky position probability, thus astronomers can choose to perform only the first $N$ observations if not enough telescope time is available to complete them all. The field containing the counterpart is at $p=25$ in this case. To complete the first 25 observations, 7 hours of telescope time are needed: this represents the minimum amount of telescope time for the EM counterpart to be detected by this facility in this case. 

Figure~\ref{fig:28840_obsmap_OPT} shows the positions and times of the observations scheduled by the algorithm in this case. During the first night, essentially all points of the detectability map are above the limit $\lambda$, thus observations are concentrated around the centres of the two large uncertainty regions (see Fig.~\ref{fig:28840_skymap} for an all-sky view), which are the points of largest skymap probability. During the second night, the detectability has fallen below the limit in most of the central parts of the two uncertainty regions, thus the algorithm moves towards points of lower skymap probability, but higher detectability. The evolution of the detectability map proceeds in a similar fashion until the fifth night, when the detectability at all points of the skymap eventually falls below $\lambda$. 

\subsection{Radio search}

\begin{figure}
 \includegraphics[width=\columnwidth]{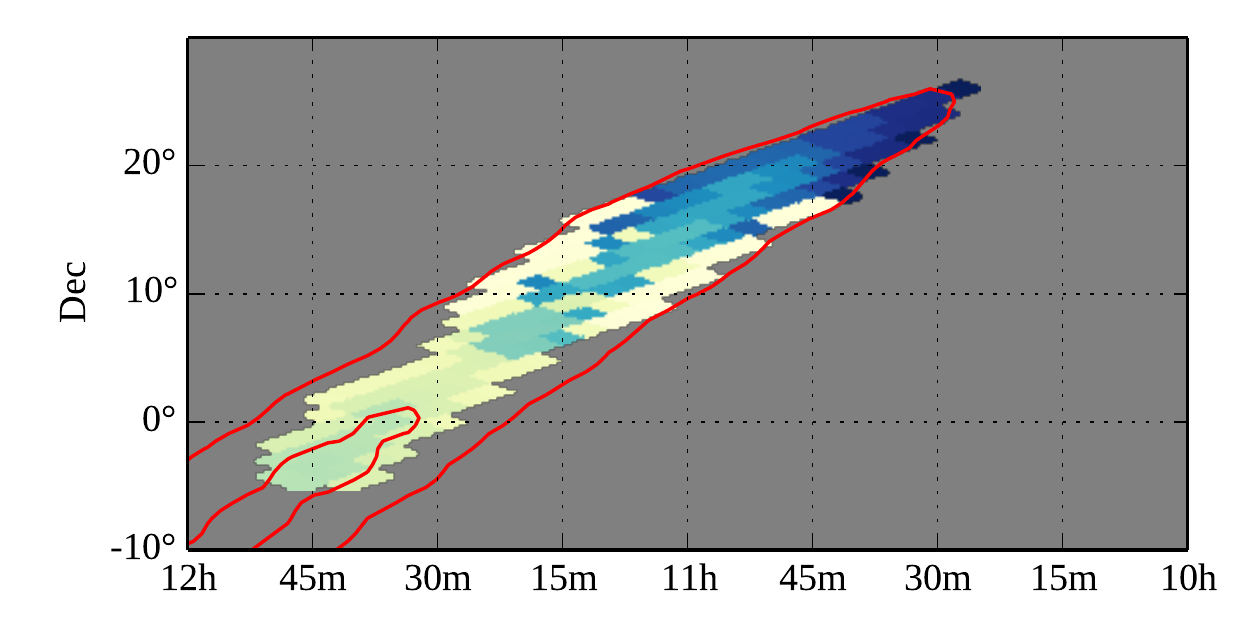}
 
 \includegraphics[width=\columnwidth]{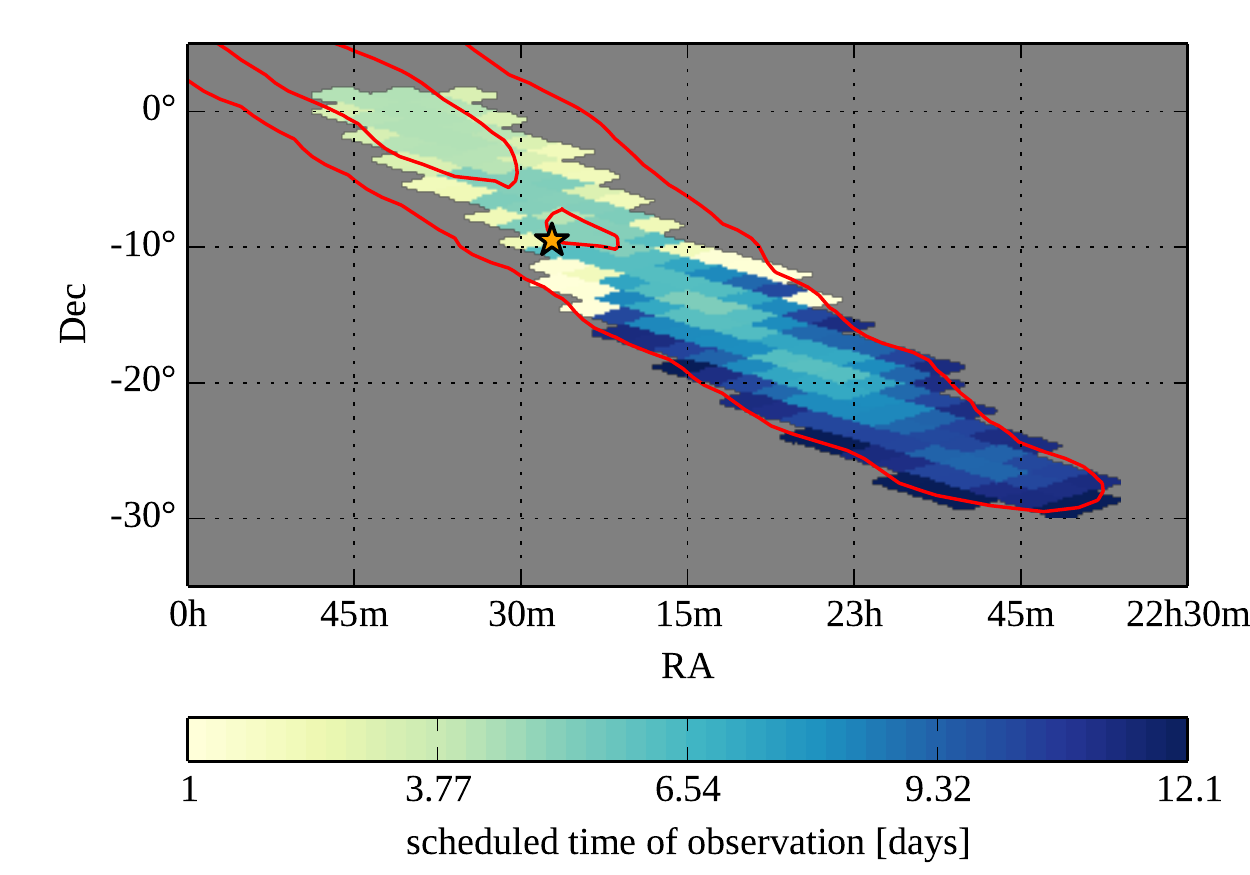}
 
 \caption{\label{fig:28840_obsmap} Time and position of the Radio follow-up observations of our test example (\S\ref{sec:example}). The star marks the injection position.}
\end{figure}

The parameters of our virtual radio follow-up are inspired by MeerKAT, the South African SKA precursor. Sixteen (of the eventual 64) 13.5 m dishes have already been integrated into a working radio telescope and produced their ``first light'' image\footnote{Media release at \url{http://www.ska.ac.za/media-releases/}} in July 2016. We assume a field of view $A\us{FoV}=1.7$ deg$^2$ at $\nu\us{obs}=1.4$ GHz. We conservatively estimate 50 $\mu$Jy rms noise for a $T\us{int}=1000$ s observation (slew + integration), assuming 16 working dishes. In a large area survey, usually a 10 sigma detection is required to avoid a large number of false alarms, thus we set $\Flim = 0.5$ mJy, \ie we require the flux to be ten times the rms noise for the detection to be considered confident. We allow a maximum of 20 percent of the available time from day 1 to day 100 to be dedicated to the follow-up (in practice, we allow an available time window of 4.8 hours each day), and we adopt a $\lambda=0.05$ detectability limit. The best detection time map with the chosen parameters is shown in Figure~\ref{fig:28840_skymap}, where the blue contours represent regions where the detectability reaches the required detectability limit at some time $t>1$ d (382 deg$^2$ in total). Setting $N\us{side}=64$, the follow-up construction algorithm outputs 337 ``pseudo observations'' (corresponding to 169 pointings, totalling 47 hours of telescope time), which are represented in Figure~\ref{fig:28840_obsmap}. The $p=35$ observation contains the counterpart. It is scheduled $5\mathrm{d}\,0\mathrm{h}\,9\mathrm{min}\,27\mathrm{s}$ after the event, when the flux of the EM counterpart is 0.52 mJy (see Fig.~\ref{fig:28840_lightcurves}), which means that the afterglow is detected at better than 10 sigma. 
The first 35 (pseudo) observations make up 4.7 hours of telescope time: only this amount needs to be actually allocated for the follow-up to successfully detect the radio counterpart.

\subsection{Infrared search of the associated macronova}

\begin{figure}
 \includegraphics[width=\columnwidth]{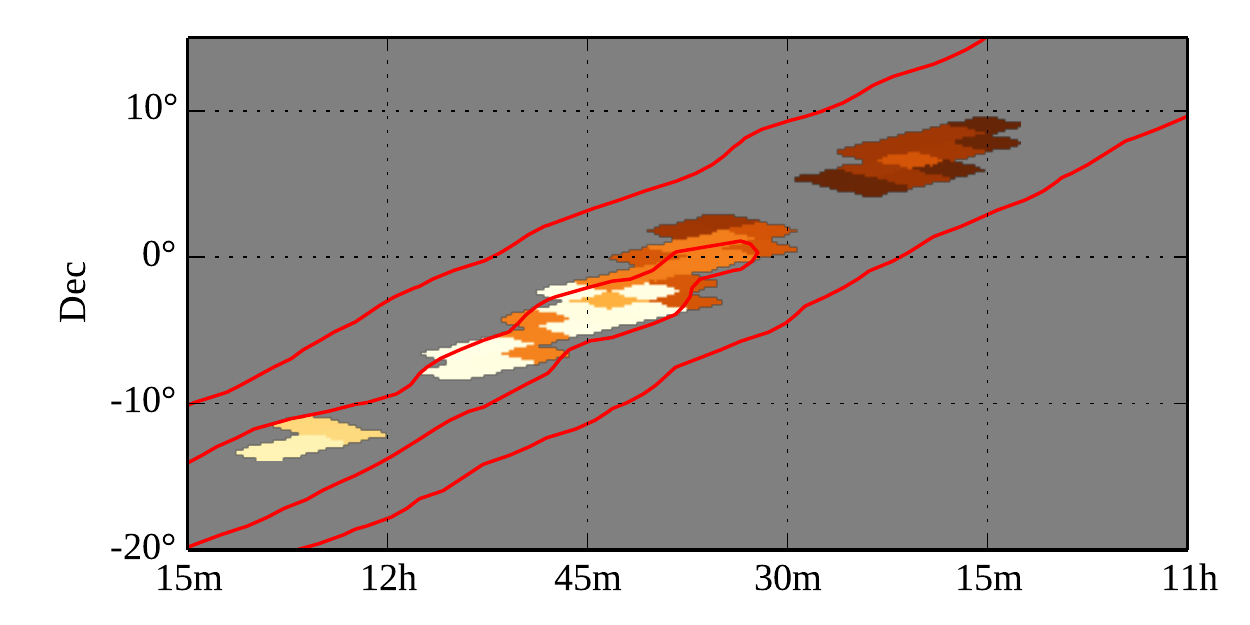}
 
 \includegraphics[width=\columnwidth]{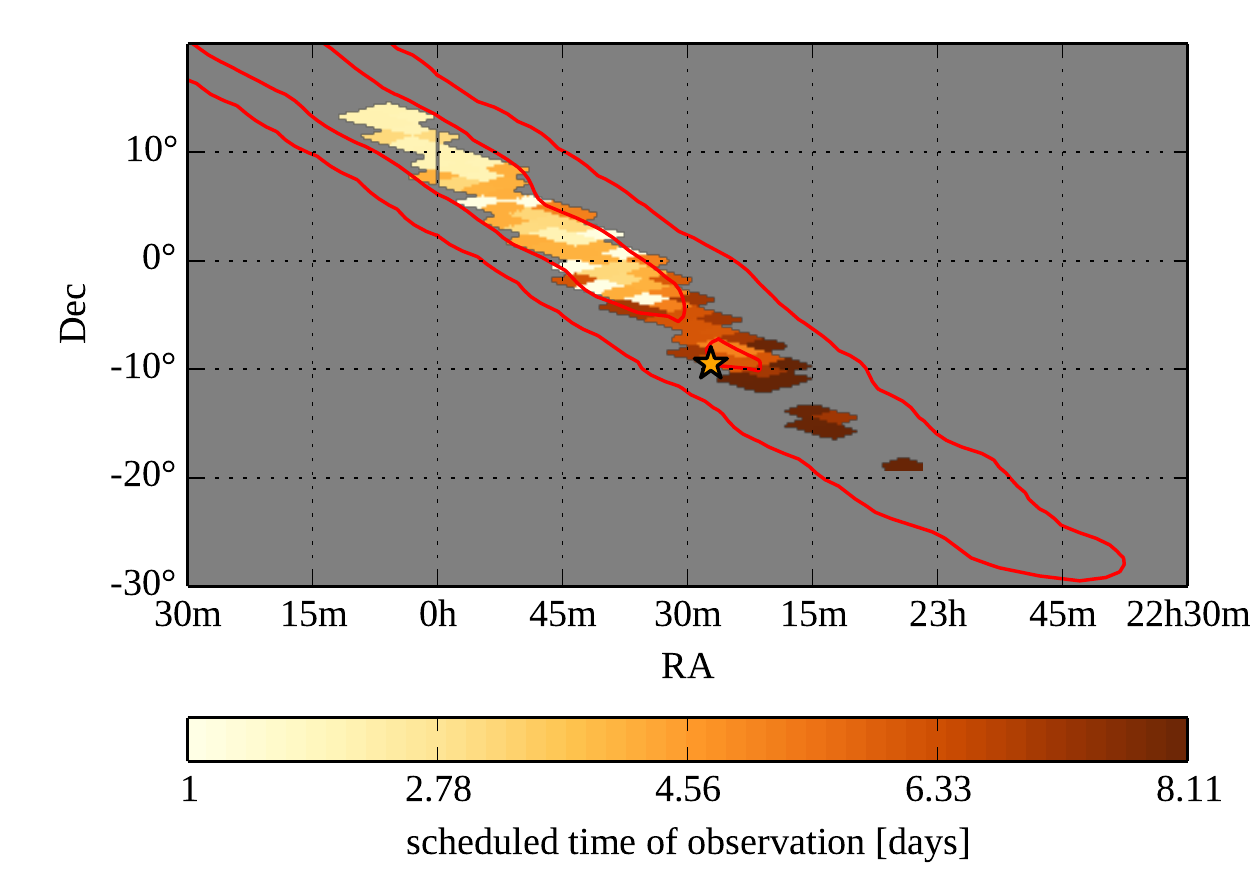}
 
 \caption{\label{fig:28840_obsmap_IR} Time and position of the Infrared follow-up observations of our test example (\S\ref{sec:example}). The star marks the injection position.}
\end{figure}

For the same event, we computed the Infrared (\textit{J} band) lightcurve (see Figure~\ref{fig:28840_lightcurves}) of the associated macronova with the same assumptions as in \S\ref{sec:a_priori_detectability_construction}. Due to the rather large masses of the binary components and to their similar mass, the dynamically ejected mass is small ($M\us{ej}\approx 5.6\times 10^{-3} M_\odot$ according to the \citet{Dietrich2016} fitting formula assuming the H4 equation of state). The lightcurve is thus quite dim. It peaks between the second and the third day, slightly brighter than 22.4 m$\us{AB}$ in the \textit{J} band. The effective temperature of the photosphere at peak is $T\us{peak}\sim 2900$ K. 
To detect such a transient with a telescope like VISTA, an integration time of the order of 1000 s is needed. We thus perform our virtual follow up strategy with the following parameters inspired to VISTA: we choose a limiting flux $\Flim=22.4$ m$\us{AB}$ in the \textit{J} band with $T\us{int}=1200$ s, and we set $A\us{FoV}=1.5$ deg$^2$. Again, we assume that 3 hours  per night are dedicated to the follow-up. The a priori detectability (see Figure~\ref{fig:PF_gt_Flim_AP}) of the macronova for this limiting flux is high, meaning that most of the possible lightcurves exceed 22.4 m$_\mathrm{AB}$, thus we set $\lambda=0.5$ to limit the search to points of the skymap which reach a detectability at least as good as the a priori one. 

Adopting the above parameters, the algorithm outputs 123 ``pseudo observations'' (corresponding to 69 pointings, totalling 23 hours of telescope time), which are represented in Figure~\ref{fig:28840_obsmap_IR}. The 79th observation in descending order of $P_p$ contains the counterpart. It is scheduled 6d 0h 18min 27s after the event, when the flux of the macronova in the \textit{J} band is just below the 22.9 m$\us{AB}$ and the effective temperature of the spectrum is around 2600 K. The flux is lower than $\Flim$ (which was chosen to represent an indicative limit for achieving a S/N ratio $\sim$ 10), but according to ESO's exposure time calculator\footnote{We queried the VIRCAM ETC at \url{http://www.eso.org} assuming a 1.2 airmass and a seeing of 0.8 arcsec.}, such emission would be detected through the VISTA telescope at ESO in the \textit{J} band with a S/N ratio of 6, assuming a 1000 s integration, in optimal observing conditions. Our virtual Infrared follow-up would thus again result in a detection in a search with a threshold at S/N $\sim$ 5. To achieve it, 14.5 hours of telescope time are needed (the amount of time for the first 79 most important observations to be performed).

\subsection{Comparison with follow-up strategies based on the a priori detectability only}
For comparison, we performed additional Optical, Radio and Infrared searches using the same parameters as before (listed in Table~\ref{tab:algorithm_inputs}), but replacing the a posteriori detectability $P(F(t) > \Flim\,|\,\alpha,\Sgnl)$ with the a priori detectability $P(F(t) > \Flim)$ computed in \S\ref{sec:a_priori_detectability_construction}. This should simulate a search based on a priori information only. The results are the following:
\begin{itemize}
 \item Optical search: the counterpart is in the field of view 5d 22h 53m 32s after the event, when the flux of the jet afterglow in the \textit{r} band is as low as 26 m$\us{AB}$, which is definitely too faint for a detection;
 \item Radio search: the counterpart is in the field of view 20d 2h 18m 0s after the event, when the flux of the jet afterglow at 1.4 GHz is 120 $\mu$Jy, which is significantly below our required limiting flux. Even assuming that the sensitivity was good enough for a detection, the facility should have allocated at least 77.6 hours of telescope time for this single follow-up in order to include the observation that contains the counterpart;
 \item Infrared search: none of the 122 fields of view whose observation is scheduled by the algorithm contains the counterpart.
\end{itemize}

We conclude that the use of posterior information from the GW signal has a decisive impact on the EM follow-up in our example test case.

\section{Discussion}\label{sec:discussion}

With this work, we proposed the new idea that information on compact binary inspiral parameters extracted from a GW signal can be used to predict (to some extent) the best timing for observation of the possible EM counterpart. In practice, the probability distributions of the binary parameters inferred from the GW signal are fed to a model of the candidate EM counterpart in order to define a family of possible lightcurves. The possible lightcurves are then used to construct the ``detectability maps'', which represent an estimate of how likely is the detection of the EM counterpart with a given instrument, if the observation is performed at time $t$ looking at sky position $\alpha=(\mathrm{RA}, \mathrm{Dec})$. In order to apply the idea to a practical example, we introduced an explicit method to construct the detectability maps (\S\ref{sec:idw_based_method}) and an algorithm which uses these maps to define an EM follow-up observing strategy (\S\ref{sec:algorithm}). We then applied the method to a synthetic example, showing that it improves significantly the effectiveness of the EM follow-up (\S\ref{sec:example}).
 
In order to keep the treatment as simple as possible, we adopted many simplifications at various stages of the discussion, and we intentionally avoided to mention some secondary details or to include too much complexity. Let us briefly address some of the points that were not discussed in the preceding sections.

\subsection{Model dependence and inclusion of priors on unknown parameters}
The approach clearly relies on the availability of models of the EM counterparts, and on our confidence in the predictions of these models. On the other hand, since the lightcurves are treated in a statistical sense, the models only need to represent correctly the peak flux and lightcurve general evolution. Fine details are lost in the processing of the lightcurves, and are thus unnecessary. Moreover, it is very straightforward to include our uncertainty on model parameters unrelated to the GW signal. In all examples discussed in this work we fixed the values of such parameters (\eg the kinetic energy $E\us{K}$ and ISM density $n\us{ISM}$ of the SGRB afterglow, or the NS equation of state). A better approach (at the cost of a higher computational cost) would be to assume priors for these parameters, \ie to assign a probability distribution to the values of these parameters based on some prior information (\eg available astrophysical data, if any) or on theoretical arguments. In this case, multiple lightcurves of the counterpart must be computed for each posterior sample, using different values of the unknown parameters sampled from the assumed priors. This should be the most effective way of incorporating the uncertainty on these parameters in the computation of the detectability maps (it applies as well to the a priori detectability and to the a posteriori detectability). We will explore the effect of the inclusion of such priors in the construction of a priori detectabilities, a posteriori detectabilities and detectability maps for a range of potential EM counterpart models in a future work.

\subsection{Sky position dependence of the parameters}

In \S\ref{sec:sky_position_conditional_idea} we stated that, in general, the sky-position-conditional posterior distribution of the inspiral parameters depends on the assumed sky position. The main driver of this dependence is the sky projection of the antenna patterns (\ie the sensitivity to different polarizations) of the interferometers of the network: the distribution of distances, inclinations and mass ratios compatible with a given signal, assuming that the source is at a particular sky position, is especially constrained by what the interferometers can or can not detect if the source is at that sky position. As an intuitive example, say that a particular sky position corresponds to the maximum sensitivity of one of the detectors with respect to a particular polarization, and say that the incident GW that yields the signal picked up by that detector contains no such polarization: only combinations of parameters for which the corresponding component of the strain is smaller than the limit set by the sensitivity are admissible if that sky location is assumed. If another sky location is assumed, the constraints change accordingly. As a general trend, points of the sky where the network has a higher sensitivity will correspond to a larger average distance of the source, thus implying a lower detectability for both the macronova and the SGRB afterglow. If the sky position uncertainty region is smaller than the typical angular scale over which the antenna pattern varies, the dependence of the posterior distributions of the parameters on sky position becomes less important: the a posteriori detectability contains most of the relevant information in that case. With Advanced Virgo joining the network, in many cases the sky position uncertainty region will still extend over a few hundreds of square degrees \citep{Abbott2016}; on the other hand, better information on the two polarization states of the GW signal will be available (the two interferometers of the aLIGO network are almost anti-aligned, thus the ability of the network to distinguish between the two polarization states is rather poor as of now). In the next decades, third generation interferometers will face again the same issues about sky localization. The use of detectability maps instead of the a posteriori detectability alone is thus likely to remain useful with more advanced networks as well.

\subsection{The choice of injection 28840}

The injection event used to construct the example presented in the last section was selected among those of the F2Y study. We considered a two-detector case (LIGO only), as it leads in general to larger localization uncertainties. We looked through the list for an event which was quite distant and whose orbit inclination was sufficiently inclined for the jet to be off-axis. The 28840 injection event luminosity distance is indeed rather large ($d\us{L}=75$ Mpc), the jet is slightly off-axis ($\iota = 14^\circ$), the sky position uncertainty is large (more than $1500$ deg$^2$) and the injection position is rather far away from the maximum of the skymap probability. The latter condition makes a search based only on the a priori information particularly ineffective, because the exploration of the skymap proceeds slowly (using relatively small field instruments) from the centre of the skymap (where the skymap probability is high) to the periphery (where the source is actually located). In cases when the sky location is better reconstructed (\ie the source is closer to the point of maximum skymap probability), the improvement in the EM follow-up effectiveness thanks to the detectability maps (with respect to a strategy based only on the skymap probability and on a priori information on the EM counterpart characteristics) could be less striking. We plan to study systematically the relative improvement in a future work.

\subsection{A better strategy construction algorithm}
The algorithm (\S\ref{sec:algorithm}) used to construct the virtual EM follow-up strategy of our example is admittedly oversimplified. An algorithm suited for real application should be able to:
\begin{enumerate}
 \item take into account the actual observability constraints on all points of the skymap at a given time, \eg the setting of tiles below the horizon;
 \item use a better (instrument specific) way of dividing the sky into potentially observable fields;
 \item consider the impact of airmass, expected seeing, dust extinction, stellar density in the field and other variables on the detectability;
 \item potentially use a different integration time for each tile, in order to maximize the detection probability;
 \item avoid sequences of widely separated pointings, that would result in a waste of time in slewing.
\end{enumerate}
Some recent works already addressed, at least in part, some of the above points. \citet{Ghosh2015} discussed an algorithm for the optimization of the tiling, which is totally compatible with our approach, since the detectability maps do not set a preferred tiling. \citet{Rana2016} developed and compared some ingenious algorithms that aim at maximizing the sky position probability in the search, taking into account per-tile setting and rising times. Their approach does not account for the time evolution of the EM counterpart luminosity, though, and it can result in the paradoxical situation in which the highest probability tile is observed \textit{before} the time at which the flux is high enough for a detection. Incorporating the information from the detectability maps in their method could be the starting point for a realistic automated strategy construction algorithm based on the ideas presented in this work. For what concerns point (4) above, both \citet{Coughlin2016} and \citet{Chan2017} found that an equal integration time in all observations could be sub-optimal with respect to the EM counterpart detection probability. Their assumptions, though, are significantly different from ours: both assume a constant luminosity of the EM counterpart (\ie they ignore the time variation of the flux), and they do not use astrophysically motivated priors on luminosity (the former use a flat prior, while the latter use Jeffrey's prior $L^{-1/2}$ limited to the range span by the peak luminosities of the macronova model in \citealt{Barnes2013}). It is not straightforward to figure out if their results are applicable to our more general case as well.

As an additional caveat, we only considered the case where one epoch observation per field is enough to identify transient sources. Realistically, this is only feasible when previous images can be used as reference, or where source catalogs are complete up to the survey limit. The identification of interesting transients and the removal of those unrelated to the GW source require at least two epochs of observations, which are not taken into account in our example algorithm.

\subsection{Use in conjunction with the ``galaxy targeting'' approach}
The use of detectability maps is entirely compatible with a search based on targeting candidate host galaxies. The observation of each target galaxy would simply need to be performed as close to the corresponding best detection time (as defined in \S\ref{sec:best_detection_time}) as possible. Since choosing a target galaxy corresponds to assuming a known distance to the source, an intriguing further refinement of the present method could be to consider the posterior distribution of binary parameters conditioned on both sky position and distance. This would have a great impact especially on the binary orbit inclination. In other words, it would be possible to associate a fairly well defined binary orbit inclination to each galaxy. The consequence would be that some galaxies (typically the most distant ones) would be better candidate hosts for a SGRB afterglow with respect to others, depending on the associated binary orbit inclination. 

\subsection{Computational feasibility of the approach}
The Monte Carlo approach adopted in this work, in which ``all possible'' lightcurves of an event must be computed (at least one per posterior sample, which means around $10^4$ lightcurves per observing band -- which must be increased by one or two orders of magnitude if priors on unknown parameters are included) requires a computationally effective way to produce the lightcurves. Since the aim of the approach is to assess the detectability, rather than to fit the model to observational data, simple analytic models which capture the main features of the expected lightcurves are better than complex numerical models suited for parameter estimation. The macronova model by \citet{Grossman2013} is a good example: few minutes are sufficient to compute $10^4$ lightcurves on a laptop using this model. Parallelization is instead unavoidable\footnote{In the specific case adopted in this paper, actually, the only two free parameters were the luminosity distance $d\us{L}$ and the viewing angle $\tv$. Since the involved redshifts are very low, it was sufficient to compute a small number of base lightcurves for each observing frequency, at a fixed distance, with varying viewing angle. The actual lightcurves were then computed by interpolation of the base lightcurves over the viewing angle, after rescaling to the correct distance, ignoring the negligible change in the rest frame frequency corresponding to the observer frame frequency considered. Indeed, interpolation over a table of pre-computed lightcurves can be a general way to reduce the computational cost of the evaluation of the EM counterpart models.} when such a large number of off-axis SGRB afterglow lightcurves are to be produced using \texttt{BOXFIT} \citep{VanEerten2011}.

\subsection{Reverse-engineering: tuning the models using information from the EM counterparts}
A fascinating possible future application could be to use the detectability maps to test the underlying EM counterpart models and the assumptions on the priors: when a relatively large number of inspirals involving at least one neutron star will have been detected, it will be possible to use the detectability maps to estimate \textit{how likely the detection (or non-detection) of the corresponding EM counterparts would have been} with a particular choice of priors and adopting a particular model. This could help to tune the models in order to better predict the detectability of subsequent events and  may be an alternative way to get insights into the population properties of the EM counterparts.

\section{Conclusions}\label{sec:conclusions}
The electromagnetic follow-up of a gravitational wave events is one of the major challenges that transient astronomy will face in the next years. The large localization uncertainty regions and the relatively low expected luminosity of the candidate counterparts call for highly optimized observation strategies. The results of this work show that information from the GW signal can be used to make event-specific adjustments to the EM follow-up strategy, and that such adjustments can improve significantly the effectiveness of the search at least in some cases, as shown in the example in \S\ref{sec:example}.  Advances in the theoretical modeling of the EM counterparts (\eg in our ability to predict the amount of mass ejected during the merger) will increase the effectiveness of this approach, and are thus of great importance for the astronomy community as well.

\section*{Acknowledgements}
We wish to thank Leo Singer and Christopher Berry for making the posterior samples of the F2Y study available to us. We thank Man Leong Chan, Christopher Messenger and Zhiping Jin for pointing out some relevant references that were missing in the first draft.
MB acknowledges financial support from the Italian Ministry of Education, University and Research (MIUR) through grant FIRB 2012 RBFR12PM1F. This project has received funding from the European Union's Horizon 2020 research and innovation programme under grant agreement No 653477. We also acknowledge financial support from the UnivEarthS Labex programme at Sorbonne Paris Cité (ANR-10-LABX-0023 and ANR-11-IDEX-0005-02). [...]


{\footnotesize
\bibliographystyle{aastex}
\bibliography{/home/omsharan/Dropbox/Bibliography/library.bib}}

\appendix

\section{Testing our inverse-distance-weighting-based method of sky-position-conditional density estimation}

\subsection{Test 1: reconstruction of the position-conditional mean and standard deviation of a known distribution}
As a first test, we constructed a set of 10000 mock posterior samples whose underlying probability distribution in parameter space has a known analytic form. The parameter space is 4-dimensional, the parameters being right ascension $\mathrm{RA}$, declination $\mathrm{Dec}$, luminosity distance $d\us{L}$ and a fourth quantity $q$ with no physical meaning. The $\mathrm{RA}$, $\mathrm{Dec}$ and $\dL$ parameters are independent. The right ascension is normally distributed with mean $12$h and sigma $30$min; the declination is uniformly distributed between $-7^\circ 30^\prime$ and $+7^\circ 30^\prime$; the luminosity distance is normally distributed with mean 100 Mpc and sigma 20 Mpc; the distribution of quantity $q$ depends on right ascension: its position-conditional distribution is normal, with mean and sigma given by
\begin{equation}
 \begin{array}{l}
  \displaystyle \mu_{q}(\mathrm{RA}) = \sin^2\left[8\pi\left(\frac{\mathrm{RA}}{12\mathrm{h}}-1\right)\right]\\
  \rule{0pt}{20pt}
  \displaystyle \sigma_{q}(\mathrm{RA}) = \frac{1}{5} + \frac{2}{3}\left(\frac{\mathrm{RA}}{12\mathrm{h}}-1\right)\\
 \end{array}
\end{equation}

\begin{figure}
      \includegraphics[width=0.48\columnwidth]{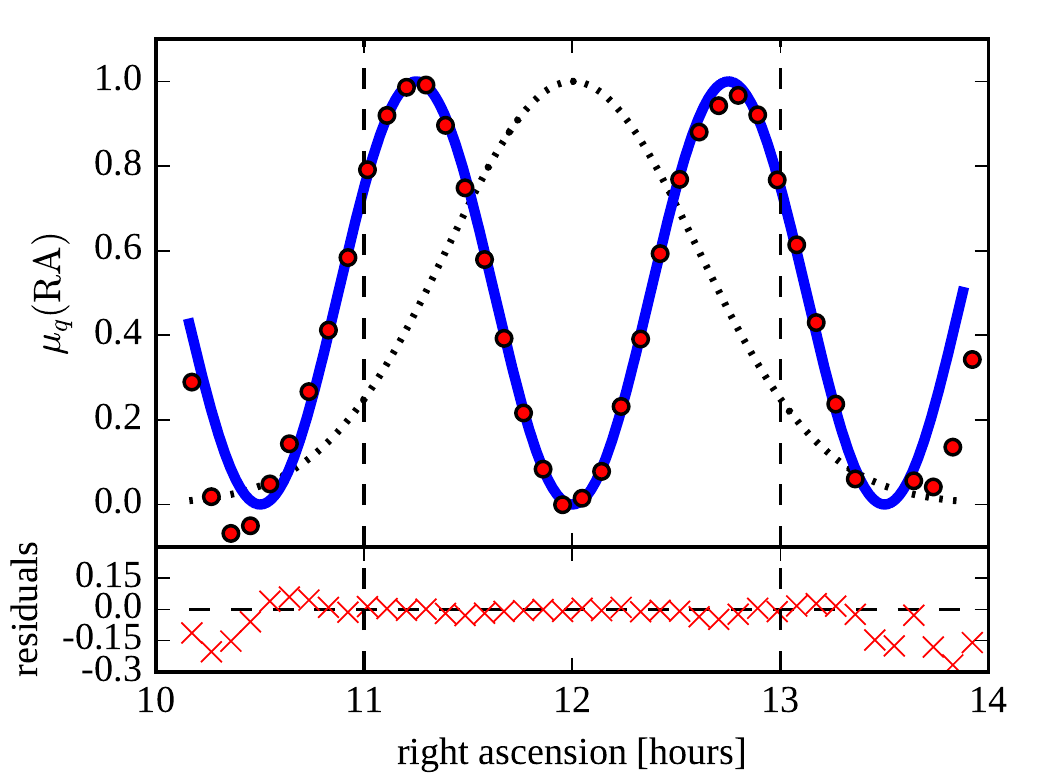}\hspace{0.03\columnwidth}\includegraphics[width=0.48\columnwidth]{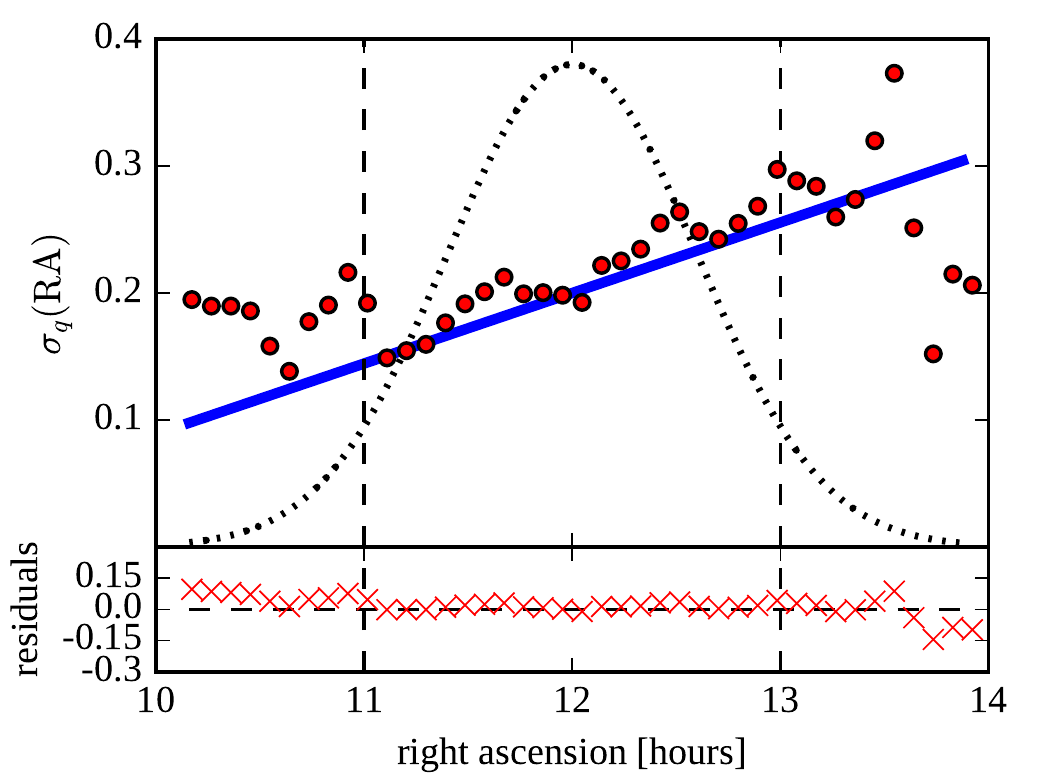}
      \caption{\label{fig:test1}Test to assess the capability of our method to recover the mean and the standard deviation of the sky-position-conditional posterior distribution of a quantity. \textbf{Left panel}: the blue line represents the position-conditional mean of the true underlying distribution of quantity $q$, while the red dots are the position-conditional means derived with our method. The black dotted line shows the density of samples around each point normalized to the maximum density, while the black dashed vertical lines show the approximate right ascension limits of the 90\% position probability area. The red crosses in the lower panel are the residuals of the computed means with respect to the true values. The accuracy in the reconstruction of the means clearly depends on the density of samples in the surrounding area. \textbf{Right panel}: same as the left panel, but for the position-conditional standard deviation. The reconstruction accuracy is clearly lower than in the case of the mean, but it remains acceptable in the region of high sample density.}

\end{figure}    

The 90\% position probability area of the posterior samples is about 435 deg$^2$ wide and its shape is approximately a rectangle extending in right ascension from 11 to 13 hours and in declination from $-7^\circ 5^\prime$ to $+7^\circ 5^\prime$. Figure \ref{fig:test1} shows the reconstructed position-conditional means and standard deviations computed in a set of points along the $\mathrm{Dec}=0^\circ$ axis. Both moments are reconstructed with an acceptable accuracy within the 90\% sky position probability area.

\subsection{Test 2: comparison with the ``Going the Distance'' study}

\begin{figure}
 \includegraphics[width=0.48\columnwidth]{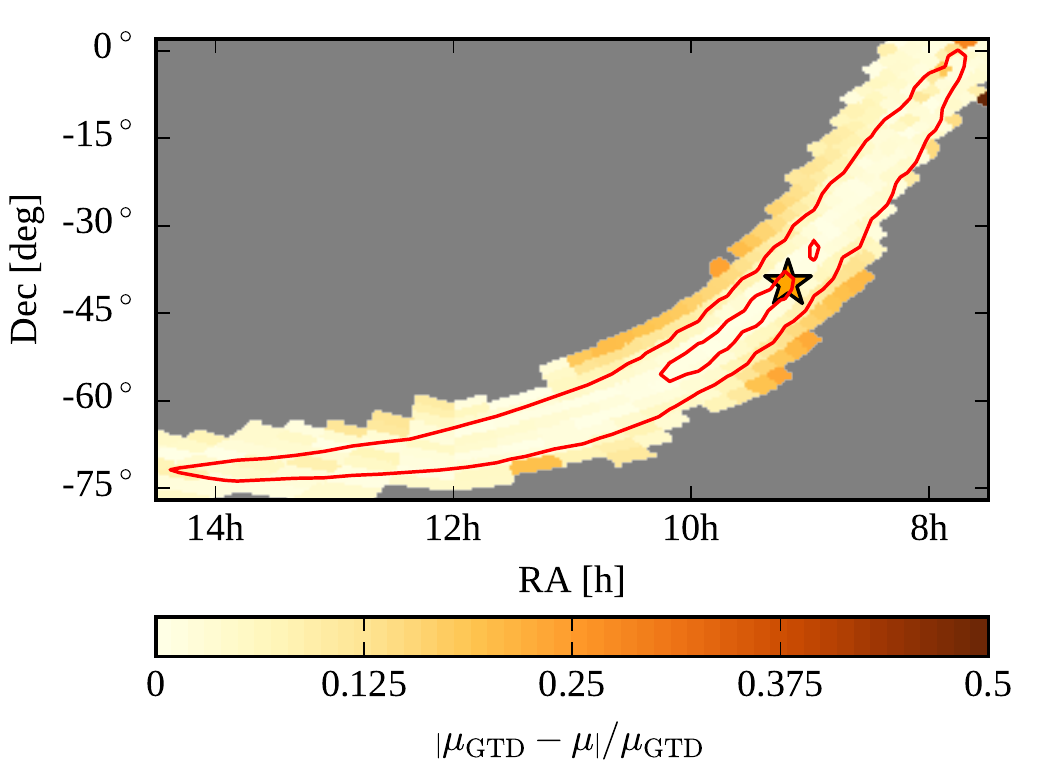}\hspace{0.03\columnwidth}\includegraphics[width=0.48\columnwidth]{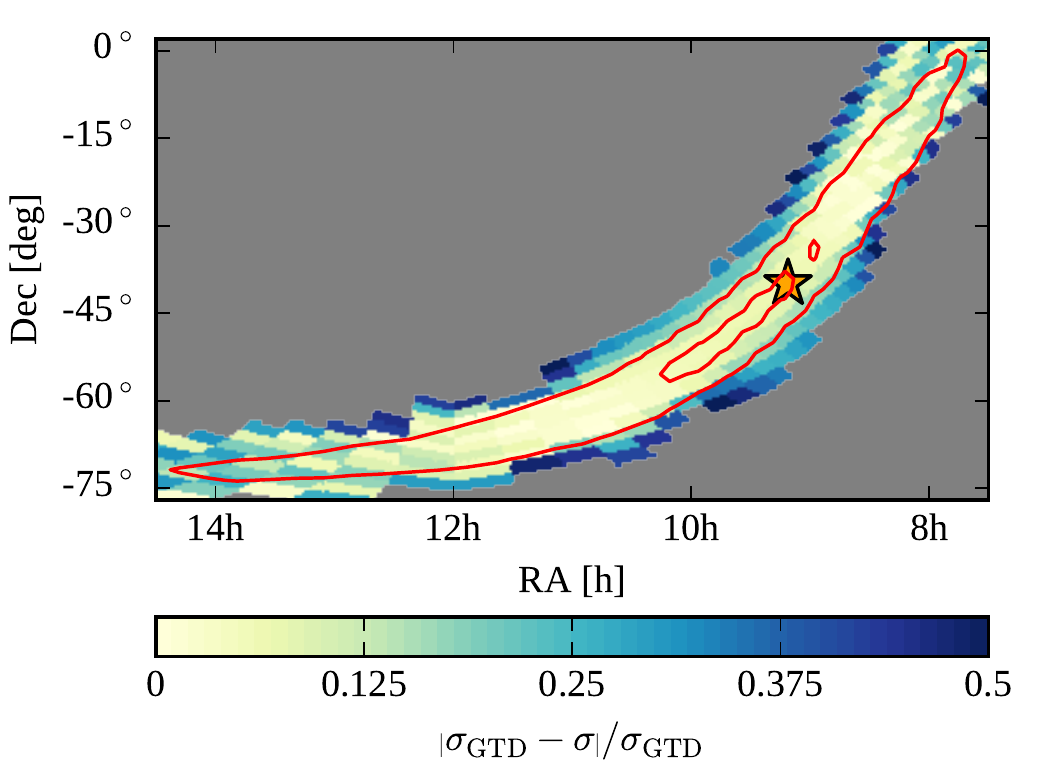}
 \caption{\label{fig:test2}Comparison between the mean of the sky-position-conditional posterior distribution of luminosity distance of injection 18951 of \citet{Singer2014} as computed with our method and that given in \citet{Singer2016a}. \textbf{Left panel:} The color coding shows the fractional deviation of the mean luminosity distance of our method compared to that of \citet{Singer2016a} (``GTD'' stands for ``Going the Distance'', \ie the title of Singer and collaborators' work). The outer (inner) red boundary represents the contour of the sky area containing 90\% (50\%) of the posterior sky position probability. The star marks the actual position of the injection. \textbf{Right panel:} Same as the left panel, but the comparison is on the standard deviation.}
\end{figure}

The ``Going the Distance'' study \citep[GTD hereafter]{Singer2016} and especially the related Supplement \citep{Singer2016a} represent an important practical step towards the use of posterior distributions of parameters other than the sky position to inform and improve the electromagnetic follow-up. In their approach, distance information encoded in the signal is used in conjunction with galaxy catalogues as the basis for a follow-up strategy based on pointing candidate host galaxies to maximise the counterpart detection probability. In the Supplement, Singer and collaborators show a step-by-step procedure to download and visualize the sky-position-conditional posterior distribution of the luminosity distance of injection 18951 from the F2Y study. We took that procedure as a starting point, and used it to compare the sky-position-conditional mean and standard deviation of luminosity distance derived with our method to those of the GTD study. Figure \ref{fig:test2} shows the relative difference between the quantities computed with the two methods. Again, the difference is very small except for regions where the density of posterior samples is small, \ie at the borders and outside the 90\% sky-position confidence region.

\end{document}